\documentclass[11pt]{article}
\usepackage{graphicx,epic,amscd,amssymb,amsmath}
\usepackage[all]{xy}
\usepackage[totalwidth=480pt,totalheight=680pt]{geometry}

\newcommand{\qed}{\hfill $\Box$}
\def\C{{\mathbb C}}
\def\Z{{\mathbb Z}}
\def\P{{\mathbb P}}
\def\Q{{\mathbb Q}}
\def\T{{\mathbb T}}
\def\A{{\mathbb A}}
\def\nn{{\nonumber}}
\def\bx{\mbox{\boldmath $x$}}
\def\by{\mbox{\boldmath $y$}}
\newtheorem{theorem}{Theorem}
\newtheorem{corollary}{Corollary}
\newtheorem{remark}{Remark}

\newtheorem{proposition}{Proposition}

\begin{document}

\title{\bf Birational maps conjugate to the rank 2 cluster mutations of affine types and their geometry}

\author{Atsushi Nobe\\
\small{Department of Mathematics, Faculty of Education, Chiba University,}\\
\small{1-33 Yayoi-cho Inage-ku, Chiba 263-8522, Japan}\\
\small{nobe@faculty.chiba-u.jp}}
\date{}
\maketitle

\begin{abstract}
Mutations of the cluster variables generating the cluster algebra of type $A^{(2)}_2$ reduce to a two-dimensional discrete integrable system given by a quartic birational map.
The invariant curve of the map is a singular quartic curve, and its resolution of the singularity induces a discrete integrable system on a conic governed by a cubic birational map conjugate to the cluster mutations of type $A^{(2)}_2$.
Moreover, it is shown that the conic is also the invariant curve of the quadratic birational map arising from the cluster mutations of type $A^{(1)}_1$ and the two birational maps on the conic are commutative.
Finally, the commutative birational maps are reduced as singular limits of additions of points on an elliptic curve arising as the spectral curve of the discrete Toda lattice of type $A^{(1)}_1$.
\end{abstract}

%-------------------------------------%
%-------------- SECTION --------------%
%-------------------------------------%
\section{Introduction}
Mutations in cluster algebras, which produce new cluster variables from old ones through their birational relationships called the exchange relations, can be regarded as time evolutions of discrete dynamical systems governed by birational maps.
Appropriate choices of the directions of mutations in adequate cluster algebras lead to proper discrete dynamical systems; discrete integrable systems.
In fact, since the introduction of cluster algebras by Fomin and Zelevinsky in 2002 \cite{FZ02} we have found many applications of cluster algebras in the field of discrete or quantum integrable systems such as discrete soliton equations, integrable maps on algebraic curves, discrete Painlev\'e equations and $Y$-systems \cite{FZ03-2,IIKNS10,IIKKN13,IIKKN13-2,Mase13,Okubo13,Okubo15,Mase16,Nobe16,BGM17}.

The number of cluster variables in the initial seed of a cluster algebra is called the rank of the algebra.
Since cluster algebras of rank 1 are trivial, the non-trivial lowest case is of rank 2.
There are three types of cluster algebra of rank 2; finite, affine and indeterminate types. 
Note that the type of a cluster algebra is defined to be the type of the Cartan counterparts of the exchange matrices appearing in the cluster pattern \cite{FZ02}.
Mutations in a cluster algebra of rank 2 lead to a two-dimensional discrete integrable system on a plane curve, as will be shown later.
Among discrete integrable systems, a paradigmatic family of  two-dimensional integrable maps called the QRT maps plays a crucial role \cite{QRT89,Tsuda04,Duistermaat10}.
For example, many reductions of discrete soliton equations and many autonomous limits of discrete Painlev\'e equations are members of the family of QRT maps. 
Moreover, since QRT maps are generated by point additions on elliptic surfaces, a connection with a QRT map suggests a geometric aspect of the integrable system.
Therefore, it seems important to investigate the cluster algebras of rank 2 thoroughly in order to grasp integrable structures of cluster algebras of higher rank.

In the preceding paper \cite{Nobe16}, we left the first footprint of the investigation for mutations in cluster algebras of rank 2 from the viewpoint of discrete integrable systems on plane curves, and a direct connection between the cluster algebra and the discrete Toda lattice both of which are of type $A^{(1)}_1$ was established.
In this and forthcoming papers \cite{Nobe18}, we will complete classification of the rank 2 cluster algebras of finite and affine types from the viewpoint of discrete integrable systems on plane curves.
First, in this paper, we consider rank 2 cluster algebras of affine types, namely, of types $A^{(1)}_1$ and $A^{(2)}_2$.
We reduce birational maps governing discrete integrable systems on plane curves from the mutations in these cluster algebras.
The invariant curve for the mutations of type $A^{(1)}_1$ is a conic, while the one for the mutations of type $A^{(2)}_2$ is a singular quartic curve.
The conic as the invariant curve for the mutations of type $A^{(1)}_1$ is also obtained by resolving the singularity of the invariant curve for the mutations of type $A^{(2)}_2$ in terms of its blowing-up.
Moreover, a cubic birational map on the conic, which is conjugate to the mutations of type $A^{(2)}_2$ with respect to the blowing-up, is also obtained.
We show that the map conjugate to the mutations of type $A^{(2)}_2$ and the one arising from the mutations of typ $A^{(1)}_1$ are commutative on the conic.
The commutativity of the maps on the conic is reduced from the additive group structure of an elliptic curve in a singular limit.
Since the additive group structure of the elliptic curve also leads to time evolutions of the discrete Toda lattice of type $A^{(1)}_1$, a direct connection between the rank 2 cluster algebras of affine types and the Toda lattice is established.
The rank 2 cluster algebras of finite type will be investigated in \cite{Nobe18}.

We briefly review a portion of cluster algebras \cite{FZ02,FZ03,FZ06}.
Let $\bx=(x_1,x_2,\ldots,x_n)$ be the set of generators of the ambient field $\mathcal{F}=\mathbb{QP}(\bx)$, where $\P=\left(\P,\cdot,\oplus\right)$ is a semifield endowed with multiplication $\cdot$ and auxiliary addition $\oplus$ and $\mathbb{QP}$ is the group ring of $\P$ over $\Q$.
Also let $\by=(y_1,y_2,\ldots,y_n)$ be an $n$-tuple in $\P^n$ and $B=(b_{ij})$ be an $n\times n$ skew-symmetrizable integral matrix.
The triple $(\bx,\by,B)$ is referred as the seed.
We also refer to $\bx$ as the cluster of the seed, to $\by$ as the coefficient tuple and to $B$ as the exchange matrix.
Each elements of $\bx$ and $\by$ are called a cluster variable and a coefficient, respectively.

We introduce mutations.
Let $k\in[1,n]$ be an integer.
The mutation $\mu_k$ in the direction $k$ transforms $(\bx,\by,B)$ into the seed $\mu_k(\bx,\by,B)=:(\bx^\prime,\by^\prime,B^\prime)$ defined as follows
\begin{align}
b_{ij}^\prime
&=
\begin{cases}
-b_{ij}&\mbox{$i=k$ or $j=k$},\\
b_{ij}+[-b_{ik}]_+b_{kj}+b_{ik}[b_{kj}]_+&\mbox{otherwise},\\
\end{cases}
\label{eq:mutem}\\
y_j^\prime
&=
\begin{cases}
y_k^{-1}&\mbox{$j=k$},\\
y_jy_k^{[b_{kj}]_+}(y_k\oplus 1)^{-b_{kj}}&\mbox{$j\neq k$},\\
\end{cases}
\label{eq:mutcoef}\\
x_j^\prime
&=
\begin{cases}
\displaystyle\frac{y_k\prod x_i^{[b_{ik}]_+}+\prod x_i^{[-b_{ik}]_+}}{(y_k\oplus 1)x_k}&\mbox{$j=k$},\\
x_j&\mbox{$j\neq k$},\\
\end{cases}
\label{eq:mutcv}
\end{align}
where we define $[a]_+:=\max(a,0)$ for $a\in\Z$.

Let $\T_n$ be the $n$-regular tree whose edges are labeled by $1, 2, \ldots, n$ so that the $n$ edges emanating from each vertex receive different labels.
We write $t\ \overset{k}{\begin{xy}\ar @{-}(10,0)\end{xy}}\ t^\prime$ to indicate that vertices $t,t^\prime\in\T_n$ are joined by an edge labeled by $k$.
We assign a seed $\Sigma_t=(\bx_t,\by_t,B_t)$ to every vertex $t\in\T_n$ so that the seeds assigned to the endpoints of any edge $t\ \overset{k}{\begin{xy}\ar @{-}(10,0)\end{xy}}\ t^\prime$ are obtained from each other by the mutation in direction $k$.
We refer the assignment $\T_n\ni t\mapsto\Sigma_t$ to a cluster pattern.
We write the elements of $\Sigma_t$ as follows
\begin{align*}
\bx_t=(x_{1;t},\ldots,x_{n;t}),\quad
\by_t=(y_{1;t},\ldots,y_{n;t}),\quad
B_t=(b_{ij}^t).
\end{align*}
Given a cluster pattern $\T_n\ni t\mapsto\Sigma_t$, we denote the union of clusters of all seeds in the pattern by
\begin{align*}
\mathcal{X}
=
\bigcup_{t\in\T_n}\bx_t
=
\left\{
x_{i;t}\ |\ t\in\T_n,\ 1\leq i\leq n
\right\}.
\end{align*}
The cluster algebra $\mathcal{A}$ associated with a given cluster pattern is the $\mathbb{ZP}$-subalgebra of the ambient field $\mathcal{F}$ generated by all cluster variables: $\mathcal{A}=\mathbb{ZP}[\mathcal{X}]$.

This paper is organized as follows.
In section \ref{sec:CAQRT}, we introduce the mutations of type $A^{(2)}_2$ and reduce a discrete integrable system on the projective plane $\P^2(\C)$ given by a quartic birational map from the mutations.
The invariant curve of the map is a singular quartic curve, therefore, we resolve the singularity by blowing-up the curve and obtain a conic as the strict transform of the singular curve.
We moreover obtain a discrete integrable system on the conic governed by a cubic birational map which is conjugate to the quartic birational map with respect to the blowing-up.
The conic thus obtained is nothing but the invariant curve of the discrete integrable system arising from the mutations of type $A^{(1)}_1$.
In section \ref{sec:A11A22}, we show commutativity of the birational maps on the conic arising from the mutations of type $A^{(2)}_2$ and of type $A^{(1)}_1$ by using the flipping structures of the maps.
Note that the latter map has already been investigated precisely in the preceding paper \cite{Nobe16}, and has been revealed a direct connection with the discrete Toda lattice of type $A^{(1)}_1$.
In section \ref{sec:TLCA}, we give a geometric interpretation of the commuting birational maps in terms of the additive group structure on an elliptic curve arising as the spectral curve of the discrete Toda lattice of type $A^{(1)}_1$.
Section \ref{sec:CONCL} is devoted to concluding remarks.

%-------------------------------------%
%-------------- SECTION --------------%
%-------------------------------------%
\section{Mutations of type $\boldsymbol{A^{(2)}_2}$ and birational maps}
\label{sec:CAQRT}
\begin{figure}[htbp]
\centering
$
\xymatrix{{t_{-2}}\ar @{-}(3,-2);(13,-12)^1&&t_0&\ar @{-}(33,-2);(43,-12)^1&t_2&\ar @{-}(60,-2);(70,-12)^1&\\
\ar @{--}(-14,-12);(-4,-2)&{t_{-1}}\ar @{-}(18,-12);(28,-2)^2&\ar @{-}(46,-12);(56,-2)^2&t_1&\ar @{--}(73,-12);(83,-2)&t_3\\}
$
\caption{
The regular binary tree $\T_2$.
}
\label{fig:binarytree}
\end{figure}
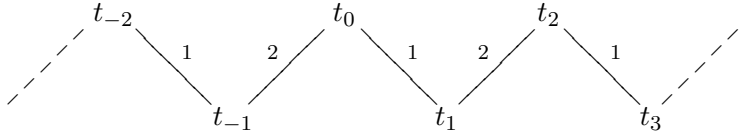

Let us consider the cluster algebra $\mathcal{A}$ generated from the following initial seed $\Sigma_0=\left(\bx_0,\by_0,B_0\right)$
\begin{align*}
\bx_0=\left(x_1,x_2\right),\quad
\by_0=\left(y_1,y_2\right),\quad
B_0=\left(\begin{matrix}0&-4\\1&0\\\end{matrix}\right),
\end{align*}
where $\bx_0$ is the cluster, $\by_0$ is the coefficient tuple and $B_0$ is the exchange matrix. 
The semifield $\P$ is arbitrarily chosen.
We consider the regular binary tree $\T_2$ whose edges are labeled by the numbers 1 and 2. 
The tree $\T_2$ is an infinite chain (see figure \ref{fig:binarytree}).
%//////////////////// FIGURE ////////////////////%
%//////////////////// FIGURE ////////////////////%
For the cluster pattern $\T_2\ni t_m\mapsto\Sigma_m=\left(\bx_m,\by_m,B_m\right)$, we denote the clusters, the coefficients and the exchange matrices by
\begin{align}
\bx_{2k}=(x_{1;2k},x_{2;2k})
&&\overset{\mu_1}{\longmapsto}&&
\bx_{2k+1}=(x_{1;2k+1},x_{2;2k+1})
&&\overset{\mu_2}{\longmapsto}&&
\bx_{2k+2}=(x_{1;2k+2},x_{2;2k+2}),
\label{eq:mu1mu2x}\\
\by_{2k}=(y_{1;2k},y_{2;2k})
&&\overset{\mu_1}{\longmapsto}&&
\by_{2k+1}=(y_{1;2k+1},y_{2;2k+1})
&&\overset{\mu_2}{\longmapsto}&&
\by_{2k+2}=(y_{1;2k+2},y_{2;2k+2}),
\label{eq:mu1mu2y}\\
B_{2k}=\left(b_{ij}^{2k}\right)
&&\overset{\mu_1}{\longmapsto}&&
B_{2k+1}=\left(b_{ij}^{2k+1}\right)
&&\overset{\mu_2}{\longmapsto}&&
B_{2k+2}=\left(b_{ij}^{2k+2}\right)
\label{eq:mu1mu2B}
\end{align}
for $k\in\Z$, respectively.

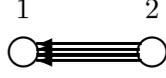
\begin{figure}[htb]
\centering
\unitlength=.075in{\def\arraystretch{1.0}
\begin{picture}(12,5)(0,0)
\thicklines
\put(1,.5){\circle{2}}
\put(10,.5){\circle{2}}
\put(9.2,1.1){\vector(-1,0){7.3}}
\put(9.1,.7){\vector(-1,0){7.2}}
\put(9.1,.3){\vector(-1,0){7.2}}
\put(9.2,-.1){\vector(-1,0){7.3}}
\put(1,3.5){\makebox(0,0){$1$}}
\put(10,3.5){\makebox(0,0){$2$}}
\end{picture}}
\caption{
The Dynkin diagram of type $A^{(2)}_2$.
}
\label{fig:Dynkin}
\end{figure}

We see from (\ref{eq:mutem}) that the exchange matrices have period two:
\begin{align*}
B_m
=
\begin{cases}
B_0&\mbox{$m$ even,}\\
-B_0&\mbox{$m$ odd.}\\
\end{cases}
\end{align*}
It follows that we have the Cartan counterpart $A(B_m)$ of $B_m$ as follows \cite{FZ03}
\begin{align*}
A(B_m)
:=
\left(2\delta_{ij}-\left|b_{ij}^m\right|\right)
=
\left(\begin{matrix}2&-4\\-1&2\\\end{matrix}\right)
\qquad
\mbox{for ${}^\forall m\in\Z$}.
\end{align*}
Since the Cartan counterparts of all the exchange matrices are the same and are of type $A^{(2)}_2$, we refer to the cluster algebra $\mathcal{A}$ as of type $A^{(2)}_2$.
The mutations $\mu_1$ and $\mu_2$ are also referred as of type $A^{(2)}_2$.
The Dynkin diagram of type $A^{(2)}_2$ is given in figure \ref{fig:Dynkin}.
%//////////////////// FIGURE ////////////////////%
%//////////////////// FIGURE ////////////////////%
Note that there is no quiver representation of the exchange matrix $B_m$ since it is not skew-symmetric but skew-symmetrizable \cite{FZ02}.

The cluster pattern $\T_2\ni t_m\mapsto\Sigma_m=\left(\bx_m,\by_m,B_m\right)$ of the cluster algebra $\mathcal{A}$ of type $A^{(2)}_2$ induces a dynamical system governed by a quartic birational map on the projective plane $\P^2(\C)$.
%//////////////////// THEOREM ////////////////////%
%//////////////////// THEOREM ////////////////////%
%//////////////////// THEOREM ////////////////////%
\begin{proposition}
Let $\mathcal{A}$ be the cluster algebra of type $A^{(2)}_2$.
For $t\geq0$, we associate $z^t$ and $w^t$ with the seed of $\mathcal{A}$ as
\begin{align*}
z^t
=
\frac{x_{1;2t}}{\sqrt[4]{y_{2;2t}}},
\qquad
w^t
=
y_{1;2t}x_{2;2t}.
\end{align*}
Assume $z^0,w^0\in\P^2(\C)$.
Then the sequence $\mu_1,\mu_2,\mu_1,\mu_2,\ldots$ of mutations in $\mathcal{A}$ induces a birational map $\psi_{\rm vh}:(z^t,w^t)\mapsto(z^{t+1},w^{t+1})$ on $\P^2(\C)$ from the initial seed $\Sigma_0=\left(\bx_0,\by_0,B\right)$, where  $z^{t+1}$ and $w^{t+1}$ are defined to be
\begin{align*}
z^{t+1}
=
\frac{w^t+1}{z^t}
\qquad
\mbox{and}
\qquad
w^{t+1}
=
\frac{\left(z^{t+1}\right)^4+1}{w^t}
\end{align*}
by using $z^t$ and $w^t$, respectively.
\end{proposition}
%//////////////////// THEOREM ////////////////////%
%//////////////////// THEOREM ////////////////////%
%//////////////////// THEOREM ////////////////////%

%---------- PROOF ----------%
(Proof)\quad
Let $\T_2\ni t_m\mapsto\Sigma_m=\left(\bx_m,\by_m,B_m\right)$ be the cluster pattern of the cluster algebra $\mathcal{A}$ of type $A^{(2)}_2$.
From the exchange relation (\ref{eq:mutcoef}) of the coefficients, it immediately follows the equalities among them:
\begin{align*}
&y_{1;2k}y_{1;2k+1}=1,
\\
&y_{2;2k+1}
=
y_{2;2k}y_{1;2k}^{[-4]_+}(y_{1;2k}\oplus 1)^4
=
y_{2;2k}(y_{1;2k}\oplus 1)^4,
\\
&y_{2;2k+1}y_{2;2k+2}=1,
\\
&y_{1;2k+2}
=
y_{1;2k+1}y_{2;2k+1}^{[-1]_+}(y_{2;2k+1}\oplus 1)^1
=
y_{1;2k+1}(y_{2;2k+1}\oplus 1).
\end{align*}
Moreover, by using the exchange relation (\ref{eq:mutcv}) of the cluster variables and the above equalities in the coefficients, we compute
\begin{align}
x_{1;2k+1}
&=
\frac{y_{1;2k}x_{2;2k}+1}{(y_{1;2k}\oplus 1)x_{1;2k}}
=
\frac{y_{1;2k}x_{2;2k}+1}{\sqrt[4]{\frac{1}{y_{2;2k}y_{2;2k+2}}}x_{1;2k}},
\nn\\
x_{2;2k+1}
&=x_{2;2k},
\nn\\
x_{1;2k+2}
&=
x_{1;2k+1}
=
\frac{y_{1;2k}x_{2;2k}+1}{\sqrt[4]{\frac{1}{y_{2;2k}y_{2;2k+2}}}x_{1;2k}},
\label{eq:x1te}\\
x_{2;2k+2}
&=
\frac{y_{2;2k+1}(x_{1;2k+1})^4+1}{(y_{2;2k+1}\oplus 1)x_{2;2k+1}}
=
\frac{\frac{1}{y_{2;2k+2}}(x_{1;2k+2})^4+1}{y_{1;2k}y_{1;2k+2}x_{2;2k}}.
\label{eq:x2te}
\end{align}

By setting
\begin{align*}
z^t
=
\frac{x_{1;2t}}{\sqrt[4]{y_{2;2t}}},
\qquad
w^t
=
y_{1;2t}x_{2;2t},
\end{align*}
we obtain the rational map $(z^t,w^t)\mapsto(z^{t+1},w^{t+1})$;
\begin{align*}
z^{t+1}
=
\frac{w^t+1}{z^t},
\qquad
w^{t+1}
=
\frac{(z^{t+1})^4+1}{w^t}
\end{align*}
from the exchange relations (\ref{eq:x1te}) and (\ref{eq:x2te}).

The inverse $\psi_{\rm vh}^{-1}:(z^{t+1},w^{t+1})\mapsto(z^t, w^t)$ is also given by
\begin{align*}
z^t
=
\frac{w^t+1}{z^{t+1}},
\qquad
w^t
=
\frac{(z^{t+1})^4+1}{w^{t+1}}.
\end{align*}
Thus, the map $\psi_{\rm vh}$ is birational.
\qed
%---------- PROOF ----------%

We find that the birational map $\psi_{\rm vh}$ is integrable in the sense of Liouville.
Actually, the two-dimensional map $\psi_{\rm vh}$ has an invariant curve parametrized by a conserved quantity depending on the initial point $(z^0,w^0)$.
The invariant curve is concretely constructed as follows.

%-------------------------------------%
%-------------- SUBSECTION --------------%
%-------------------------------------%
\subsection{Invariant curve}
Let $\overline\gamma_\lambda$ be a curve on the projective plane $\P^2(\C)$ defined to be
\begin{align*}
&\overline\gamma_\lambda
:=
\gamma_\lambda\cup\left\{P_\infty^\prime\right\},
\\
&\gamma_\lambda
:=
\left(f(z,w)=0\right),
\\
&f(z,w):=(w+1)^2+\lambda z^2w+z^4,
\end{align*}
where the point $P_\infty^\prime$ at infinity is given by $[0:1:0]$ in the homogeneous coordinate $(z,w)\mapsto[z:w:1]$ and $\lambda\in\P^1(\C)$ is a parameter.
The curve $\gamma_\lambda$ is the affine part of $\overline\gamma_\lambda$.
The base points of the pencil $\left\{\overline\gamma_\lambda\right\}_{\lambda\in\P^1(\C)}$ are the following 6 points:
\begin{align*}
&P_\infty^\prime=[0:1:0]\quad(\mbox{with multiplicity four}),
\nn\\
&\mathfrak{p}:=[0:-1:1]\quad(\mbox{with multiplicity four}),
\\
&\left[{\zeta_8}:0:1\right],\quad
\left[{\zeta_8}^3:0:1\right],\quad
\left[{\zeta_8}^5:0:1\right],\quad
\left[{\zeta_8}^7:0:1\right],
\nn
\end{align*}
where $\zeta_8$ is the eighth root of 1.
Let $\overline{\mathcal{B}}$ and $\mathcal{B}$ be the sets of the base points:
\begin{align*}
&\overline{\mathcal{B}}
:=
\mathcal{B}
\cup
\left\{P_\infty^\prime\right\},
\\
&\mathcal{B}
:=
\left\{
\mathfrak{p},
\left[{\zeta_8}:0:1\right],
\left[{\zeta_8}^3:0:1\right],
\left[{\zeta_8}^5:0:1\right],
\left[{\zeta_8}^7:0:1\right]
\right\}.
\end{align*}

\begin{figure}[htb]
\centering
\includegraphics[scale=.75]{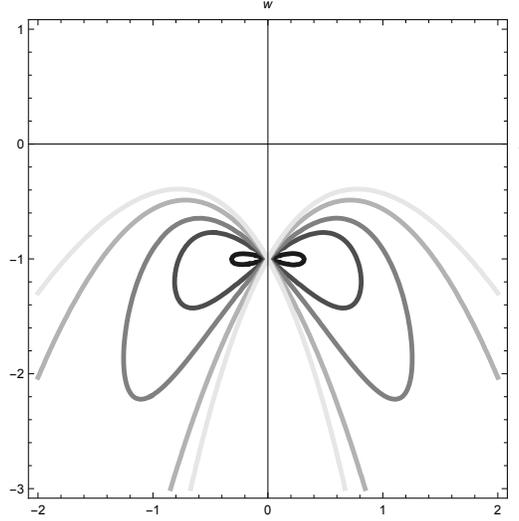}
\caption{Members of the pencil $\left\{\overline\gamma_\lambda\right\}_{\lambda\in\P^1(\C)}$ for $\lambda=0.1$, $0.6$, $1.1$, $2.1$ and $3.1$.
A curve with smaller $\lambda$ is colored darker.
Each curve passes through the base point $\mathfrak{p}$.}
\label{fig:pencil}
\end{figure}

%//////////////////// THEOREM ////////////////////%
%//////////////////// THEOREM ////////////////////%
%//////////////////// THEOREM ////////////////////%
\begin{remark}
The curve $\overline\gamma_\lambda$ is a singular quartic curve which has the singularity at the point $\mathfrak{p}$.
The singular point $\mathfrak{p}$ is an ordinary double point and is the base point of the pencil $\left\{\overline\gamma_\lambda\right\}_{\lambda\in\P^1(\C)}$ as well (see figure \ref{fig:pencil}).
\end{remark}
%//////////////////// THEOREM ////////////////////%
%//////////////////// THEOREM ////////////////////%
%//////////////////// THEOREM ////////////////////%

%//////////////////// FIGURE ////////////////////%
%//////////////////// FIGURE ////////////////////%

The singular curve $\overline\gamma_\lambda$ is nothing but the invariant curve of the map $\psi_{\rm vh}$.
%//////////////////// THEOREM ////////////////////%
%//////////////////// THEOREM ////////////////////%
%//////////////////// THEOREM ////////////////////%
\begin{theorem}\label{thm:A22ds}
Assume $(z^0,w^0)$ to be a point on $\P^2(\C)-\overline{\mathcal{B}}$.
Let $S:=\left\{(z^t,w^t)\right\}_{t\geq0}$ be a sequence of points on $\P^2(\C)$ generated from $(z^0,w^0)$ by applying  the map $\psi_{\rm vh}$ repeatedly.
Then any point in $S$ is on the affine curve $\gamma_\lambda$ for
\begin{align}
\lambda
=
-\frac{(w^0+1)^2+(z^0)^4}{(z^0)^2w^0}.
\label{eq:lambda}
\end{align}
\end{theorem}
%//////////////////// THEOREM ////////////////////%
%//////////////////// THEOREM ////////////////////%
%//////////////////// THEOREM ////////////////////%

%---------- PROOF ----------%
(Proof)\quad
Since $(z^0,w^0)$ is not a base point of the pencil $\left\{\overline\gamma_\lambda\right\}_{\lambda\in\P^1(\C)}$, it determines the value of the parameter $\lambda\in\P^1(\C)$ by (\ref{eq:lambda}), uniquely.
Note that the only point $P^\prime_\infty$ at infinity on $\overline\gamma_\lambda$ is in $\overline{\mathcal{B}}$.
Hence, we assume that $(z^t,w^t)$ is on the affine curve $\gamma_\lambda$ for $t\geq0$.
We moreover assume $(z^t,w^t)\not\in\overline{\mathcal{B}}$ by virtue of the birationality of the map $\psi_{\rm vh}$.
We then have $f(z^t,w^t)=0$, or equivalently
\begin{align*}
\lambda
=
-\frac{(w^t+1)^2+(z^t)^4}{(z^t)^2w^t}
\end{align*}
for $t\geq0$.

First we consider the horizontal flip $\psi_{\rm h}:(z^t,w^t)\mapsto(z^{t+1},w^t)$.
Through $\psi_{\rm h}$, the point $(z^t,w^t)\in\gamma_\lambda$ is mapped into the point $(z^{t+1},w^t)$.
We then have
\begin{align*}
f(z^{t+1},w^t)
&=
(w^t+1)^2-\frac{(w^t+1)^2+(z^t)^4}{(z^t)^2w^t} (z^{t+1})^2w^t+(z^{t+1})^4\\
&=
(w^t+1)^2+\left\{\left(\frac{w^t+1}{z^t}\right)^2-\frac{(w^t+1)^2+(z^t)^4}{(z^t)^2}\right\}\left(\frac{w^t+1}{z^t}\right)^2\\
&=
(w^t+1)^2-(z^t)^2\left(\frac{w^t+1}{z^t}\right)^2
=0.
\end{align*}
Thus, the point $(z^{t+1},w^t)$ is also on the curve $\gamma_\lambda$.

Next we consider the vertical flip $\psi_{\rm v}:(z^{t+1},w^t)\mapsto(z^{t+1},w^{t+1})$, and show that the point $(z^{t+1},w^{t+1})$ is also on $\gamma_\lambda$:
\begin{align*}
f(z^{t+1},w^{t+1})
&=
(w^{t+1}+1)^2-\frac{(w^t+1)^2+(z^{t+1})^4}{(z^{t+1})^2w^t} (z^{t+1})^2w^{t+1}+(z^{t+1})^4\\
&=
\frac{1}{w^t}\left\{w^t(w^{t+1}+1)^2-(w^t+1)^2w^{t+1}-(z^{t+1})^4\left(w^{t+1}-w^t\right)\right\}\\
&=
\frac{w^{t+1}-w^t}{w^t}\left\{w^tw^{t+1}-1-(z^{t+1})^4\right\}
=
0.
\end{align*}
Here we use $w^tw^{t+1}=(z^{t+1})^4+1$ and
\begin{align*}
\lambda
=
-\frac{(w^t+1)^2+(z^t)^4}{(z^t)^2w^t}
=
-\frac{(w^t+1)^2+(z^{t+1})^4}{(z^{t+1})^2w^t}.
\end{align*}
Induction on $t$ completes the proof.
\qed
%---------- PROOF ----------%

\begin{figure}[htb]
\centering
\unitlength=.03in
\def\arraystretch{1.0}
\begin{picture}(100,50)(-5,-2)
%%%%%%%%%% line %%%%%%%%%%
\qbezier(29,32)(48,55)(70,33)
\put(68,35){\vector(1,-1){3}}
\qbezier(74,29)(95,15)(75,2.3)
\put(76.5,3){\vector(-2,-1){3}}
\qbezier(28,27)(35,-5)(68,-1)
\put(65,-1.8){\vector(3,1){4}}
\thicklines
%%%%%%%%%% curve %%%%%%%%%%
\qbezier(5,25)(15,45)(40,20)
\qbezier(5,25)(2,19)(0,10)
\qbezier(0,10)(-2,0)(8,0)
\qbezier(8,0)(20,0)(40,20)
\qbezier(40,20)(65,45)(75,25)
\qbezier(75,25)(78,19)(80,10)
\qbezier(80,10)(82,0)(72,0)
\qbezier(72,0)(60,0)(40,20)
\put(-5,30){\line(1,0){90}}
\put(71.8,40){\line(0,-1){45}}
\put(71.8,30){\circle*{2.5}}
\put(28,30){\circle*{2.5}}
\put(71.8,.2){\circle*{2.5}}
%%%%%%%%%% label %%%%%%%%%%
\put(25,38){\makebox(0,0){$\psi_{\rm h}$}}
\put(85,26){\makebox(0,0){$\psi_{\rm v}$}}
\put(22,24){\makebox(0,0){$\psi_{\rm vh}$}}
%\put(55,20){\makebox(0,0){$\gamma_1$}}
\end{picture}
\caption{
The horizontal flip $\psi_{\rm h}$, the vertical flip $\psi_{\rm v}$ and their composition $\psi_{\rm vh}=\psi_{\rm v}\circ\psi_{\rm h}$ on the singular curve $\gamma_\lambda$.
These maps are arising from the mutations of type $A^{(2)}_2$.
}
\label{fig:Flips}
\end{figure}
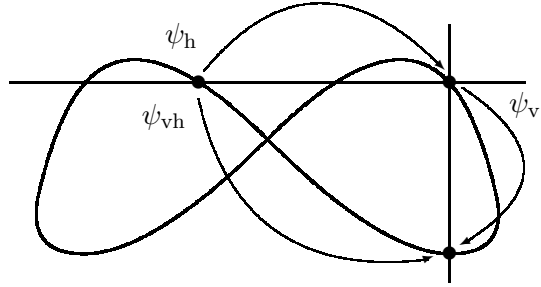

%//////////////////// THEOREM ////////////////////%
%//////////////////// THEOREM ////////////////////%
%//////////////////// THEOREM ////////////////////%
\begin{remark}
The map $\psi_{\rm vh}$ is a map of QRT type, {\it viz}, $\psi_{\rm vh}$ is the composition of the horizontal flip $\psi_{\rm h}:(z^t,w^t)\mapsto(z^{t+1},w^t)$ and the vertical flip $\psi_{\rm v}:(z^{t+1},w^t)\mapsto(z^{t+1},w^{t+1})$ (see the proof of theorem \ref{thm:A22ds} and figure \ref{fig:Flips}).
This is the reason why we denote the map by $\psi_{\rm vh}$.
We will reveal that such a flipping structure comes from the addition of points on an elliptic curve, later on.
\end{remark}
%//////////////////// THEOREM ////////////////////%
%//////////////////// THEOREM ////////////////////%
%//////////////////// THEOREM ////////////////////%

%//////////////////// FIGURE ////////////////////%
%//////////////////// FIGURE ////////////////////%

%//////////////////// THEOREM ////////////////////%
%//////////////////// THEOREM ////////////////////%
%//////////////////// THEOREM ////////////////////%
\begin{remark}
By virtue of theorem \ref{thm:A22ds}, the mutations of type $A^{(2)}_2$ induces a map on the set $\gamma_\lambda-\mathcal{B}$ which is an affine curve from which 5 points are removed.
\end{remark}
%//////////////////// THEOREM ////////////////////%
%//////////////////// THEOREM ////////////////////%
%//////////////////// THEOREM ////////////////////%

%-------------------------------------%
%-------------- SUBSECTION --------------%
%-------------------------------------%
\subsection{Resolution of singularity}
Now we consider resolution of the singularity of $\overline\gamma_\lambda$ at $\mathfrak{p}$.
Since $\mathfrak{p}$ is an ordinary double point, we can resolve the singularity by blowing-up $\overline\gamma_\lambda$ at $\mathfrak{p}$ once.
Moreover, since $\mathfrak{p}$ is the base point of the pencil $\left\{\overline\gamma_\lambda\right\}_{\lambda\in\P^1(\C)}$, the singularity of the curves in the pencil are resolved by the blowing-up, all at once.

Let $V$ be the blowing-up of $\P^2(\C)$ at $\mathfrak{p}$ and $\pi:V\to\P^2(\C)$ be the projection.
Let $\mathcal{U}\simeq\A^2$ be the affine plane containing $\mathfrak{p}$.
Note that $\overline\gamma_\lambda\cap\mathcal{U}=\gamma_\lambda$.
Then the blowing-up $\widetilde U=\widetilde{\mathcal{U}_0}\cup\widetilde{\mathcal{U}_1}\subset\P^1(\C)\times\A^2$ of $\mathcal{U}$ at $\mathfrak{p}$ is given by
\begin{align*}
&\widetilde U
:=
\left\{
\left((a_0:a_1),(b,c)\right)\ |\
ba_1=(c+1)a_0
\right\},
\\
&\widetilde{\mathcal{U}_i}
:=
\left\{
\left((a_0:a_1),(b,c)\right)\in\widetilde U\ |\
a_i\neq0
\right\}
\quad
(i=0,1).
\end{align*}
The subset $\widetilde U$ of $\P^1(\C)\times\A^2$  is obtained by patching together $\widetilde{\mathcal{U}_0}$ and $\widetilde{\mathcal{U}_1}$ as follows.

There exist isomorphisms $\iota_0:\A^2\to\widetilde{\mathcal{U}_0}$;
\begin{align*}
(u,v)\mapsto\left((1:u),(v,uv-1)\right)
\end{align*}
and $\iota_1:\A^2\to\widetilde{\mathcal{U}_1}$;
\begin{align*}
(x,y)\mapsto\left((x:1),((y+1)x,y)\right).
\end{align*}
Therefore, the composition $\iota=\iota_1^{-1}\circ\iota_0$ of these isomorphisms gives the isomorphism $\iota: \A^2-\{u=0\}\to\A^2-\{x=0\}$;
\begin{align*}
\iota:(u,v)\mapsto(x,y)=\left(\frac{1}{u},uv-1\right).
\end{align*}
By glueing $\widetilde{\mathcal{U}_0}$ with $\widetilde{\mathcal{U}_1}$ along $\iota$, we obtain $\widetilde U$.

\begin{figure}[htb]
\centering
\includegraphics[scale=.75]{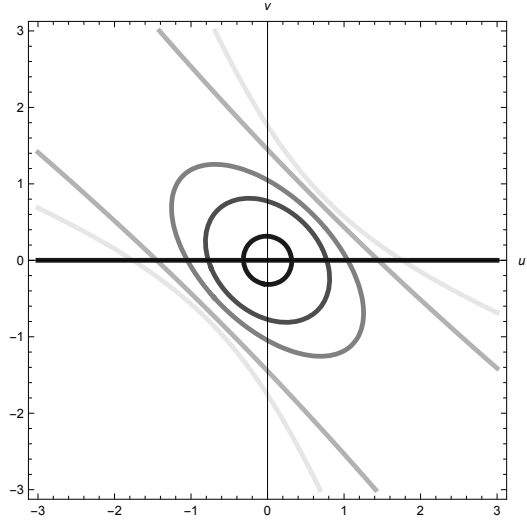}
\caption{The exceptional curve ($v=0$) and the members of the pencil $\left\{\widetilde{\gamma}_\lambda\right\}_{\lambda\in\P^1(\C)}$ of the strict transforms of $\gamma_\lambda$ for $\lambda=0.1$, $0.6$, $1.1$, $2.1$ and $3.1$.
A curve with smaller $\lambda$ is colored darker.
Each curve in $\left\{\widetilde{\gamma}_\lambda\right\}_{\lambda\in\P^1(\C)}$ intersects the exceptional curve at the points $\left(\pm\sqrt{\lambda},0\right)$.}
\label{fig:pencilresolved}
\end{figure}

On the affine plane $\widetilde{\mathcal{U}_0}\simeq\A^2$, the projection $\widetilde\pi:\widetilde U\to\mathcal{U}$ is given by
\begin{align*}
\widetilde\pi:(u,v)\mapsto (z,w)
=(v,uv-1).
\end{align*}
The total transform of $\gamma_\lambda$ in terms of the projection $\widetilde\pi$ is given by
\begin{align*}
&v^2\widetilde {f_0}(u,v)=0,
\\
&\widetilde  {f_0}(u,v)
:=
u^2+\lambda(uv-1)+v^2.
\end{align*}
Thus the strict transform, which is denoted by $\widetilde{\gamma}_\lambda$, and the exceptional curve are defined to be
\begin{align*}
\widetilde {f_0}(u,v)=0
\qquad
\mbox{and}
\qquad
v=0,
\end{align*}
respectively.
These curves intersect at two points (see figure \ref{fig:pencilresolved})
\begin{align*}
(u,v)=\left(\pm\sqrt{\lambda},0\right).
\end{align*}

Similarly, on the affine plane $\widetilde{\mathcal{U}_1}\simeq\A^2$, the projection $\widetilde\pi:\widetilde U\to\mathcal{U}$ is given by
\begin{align*}
\widetilde\pi:(x,y)\mapsto (z,w)=((y+1)x,y).
\end{align*}
The total transform of $\gamma_\lambda$ in terms of the projection $\widetilde\pi$ is given by
\begin{align*}
&(y+1)^2\widetilde {f_1}(x,y)=0,
\\
&\widetilde  {f_1}(x,y)
:=
1+\lambda x^2y+(y+1)^2x^4.
\end{align*}
Thus the strict transform $\widetilde{\gamma}_\lambda$ and the exceptional curve are defined to be
\begin{align*}
\widetilde {f_1}(u,v)=0
\qquad
\mbox{and}
\qquad
y+1=0,
\end{align*}
respectively.
These curves intersect at two points
\begin{align*}
(x,y)=\left(\pm\frac{1}{\sqrt{\lambda}},-1\right).
\end{align*}

%//////////////////// FIGURE ////////////////////%
%//////////////////// FIGURE ////////////////////%

Let us transform the map $\psi_{\rm vh}:(z^t,w^t)\mapsto(z^{t+1},w^{t+1})$ on the singular curve $\gamma_\lambda-\mathcal{B}$ into $\widetilde\psi_{\rm dv}:(u^t,v^t)\mapsto(u^{t+1},v^{t+1})$ on the non-singular curve $\widetilde{\gamma}_\lambda$ through the projection $\widetilde\pi$:
\begin{align*}
&\widetilde\psi_{\rm dv}
=
\widetilde\pi^{-1}\circ\psi_{\rm vh}\circ\widetilde\pi,
\\
&\begin{CD}
\gamma_\lambda @<\widetilde\pi<< \widetilde\gamma_\lambda
\\
@A \psi_{\rm vh} AA @AA \widetilde\psi_{\rm dv} A
\\
\gamma_\lambda @<\widetilde\pi<< \widetilde\gamma_\lambda.
\end{CD}
\end{align*}
Since $(u^t,v^t)\neq\mathfrak{p}$, we have only to consider $\widetilde{\gamma}_\lambda$ on $\widetilde{\mathcal{U}_0}$:
\begin{align*}
\widetilde{\gamma}_\lambda
=
\left(
\widetilde  {f_0}(u,v)=0
\right)
=
\left(
u^2+\lambda(uv-1)+v^2=0
\right).
\end{align*}
We then have
\begin{align}
u^{t+1}
=\frac{(u^t)^3+v^t}{u^tv^t-1},
\qquad
v^{t+1}=u^t.
\label{eq:tildepsi}
\end{align}
It immediately follows the following proposition.

%//////////////////// THEOREM ////////////////////%
%//////////////////// THEOREM ////////////////////%
%//////////////////// THEOREM ////////////////////%
\begin{proposition}
The birational map $\widetilde\psi_{\rm dv}:(u^t,v^t)\mapsto(u^{t+1},v^{t+1})$ (\ref{eq:tildepsi}) induces a map on the non-singular curve from which 4 points are removed:
\begin{align*}
&\widetilde{\gamma}_\lambda
-
\widetilde{\mathcal{B}},
\\
&\widetilde{\mathcal{B}}
:=
\left\{
(\zeta_8^{8-i},\zeta_8^i)\ |\ 
i=1,3,5,7
\right\}.
\end{align*}
Moreover, the map $\widetilde\psi_{\rm dv}$ is the composition $\widetilde\psi_{\rm d}\circ\widetilde\psi_{\rm v}$ of the vertical flip
\begin{align*}
\widetilde\psi_{\rm v}:(u^t,v^t)\mapsto\left(u^t,\frac{(u^t)^3+v^t}{u^tv^t-1}\right)
\end{align*}
and the diagonal flip
\begin{align*}
\widetilde\psi_{\rm d}:(u^t,v^t)\mapsto(v^t,u^t).
\end{align*}
\end{proposition}
%//////////////////// THEOREM ////////////////////%
%//////////////////// THEOREM ////////////////////%
%//////////////////// THEOREM ////////////////////%

%---------- PROOF ----------%
(Proof)\quad
First note that the base points $[\zeta_8^i:0:1]$ ($i=1,3,5,7$) of the pencil $\left\{\gamma_\lambda\right\}_{\lambda\in\C}$ are mapped into the base points $(u,v)=(\zeta_8^{8-i},\zeta_8^i)$ ($i=1,3,5,7$) of the pencil $\left\{\widetilde{\gamma}_\lambda\right\}_{\lambda\in\C}$ by $\widetilde\pi^{-1}$, respectively.
There exists no other base points of $\left\{\widetilde{\gamma}_\lambda\right\}_{\lambda\in\C}$.
The base point $\mathfrak{p}$ of $\left\{\gamma_\lambda\right\}_{\lambda\in\C}$ is transformed into the exceptional curve, which intersects $\widetilde{\gamma}_\lambda$ at $\left(\pm\sqrt{\lambda},0\right)$.
Since the inverse $\widetilde\pi^{-1}$ of the projection $\widetilde\pi$ is not uniquely defined at $\left(\pm\sqrt{\lambda},0\right)$, we extend $\widetilde\psi_{\rm dv}$ at $\left(\pm\sqrt{\lambda},0\right)$ by using (\ref{eq:tildepsi}).
Therefore, we assume that the initial point $(u^0,v^0)$ satisfies
\begin{align*}
(u^0,v^0)\neq(\zeta_8^{8-i},\zeta_8^i)\quad (i=1,3,5,7).
\end{align*}
Then the value of $\lambda$ is uniquely determined
\begin{align*}
\lambda
=
-\frac{(u^0)^2+(v^0)^2}{u^0v^0-1}.
\end{align*}

Assume that $(u^t,v^t)$ is on $\widetilde{\gamma}_\lambda$.
We then have $\widetilde{f}(u^t,v^t)=0$, or equivalently
\begin{align*}
\lambda
=
-\frac{(u^t)^2+(v^t)^2}{u^tv^t-1}.
\end{align*}
We show that the point $\left(u^t,\frac{(u^t)^3+v^t}{u^tv^t-1}\right)$ is also on $\widetilde{\gamma}_\lambda$:
\begin{align*}
\widetilde{f}_0\left(u^t,\frac{(u^t)^3+v^t}{u^tv^t-1}\right)
&=
(u^t)^2-\frac{(u^t)^2+(v^t)^2}{u^tv^t-1}\left(u^t\frac{(u^t)^3+v^t}{u^tv^t-1}-1\right)+\left(\frac{(u^t)^3+v^t}{u^tv^t-1}\right)^2
\\
&=
(u^t)^2-\frac{(u^t)^4(v^t)^2+(u^t)^2-2(u^t)^3v^t}{(u^tv^t-1)^2}
=0.
\end{align*}

\begin{figure}[htb]
\centering
\unitlength=.03in
\def\arraystretch{1.0}
\begin{picture}(70,70)(-33,-30)
%%%%%%%%%% line %%%%%%%%%%
\qbezier(-4,24)(15,0)(-4,-23)
\put(-1.3,-20){\vector(-1,-1){3}}
\qbezier(-8,-25)(-25,-25)(-25,-8)
\put(-25,-9){\vector(0,1){1}}
\qbezier(-8,26)(-30,30)(-26,-2)
\put(-26.5,0.5){\vector(1,-4){1}}
\thicklines
%%%%%%%%%% curve %%%%%%%%%%
\qbezier(-25,0)(-27,27)(0,25)
\qbezier(0,25)(23,23)(25,0)
\qbezier(25,0)(27,-27)(0,-25)
\qbezier(0,-25)(-23,-23)(-25,0)
\put(0,-30){\line(-1,1){30}}
\put(-6,30){\line(0,-1){60}}
\put(-6,-24.){\circle*{2.5}}
\put(-6,25){\circle*{2.5}}
\put(-24,-6){\circle*{2.5}}
%%%%%%%%%% label %%%%%%%%%%
\put(-26,-23){\makebox(0,0){$\widetilde\psi_{\rm d}$}}
\put(12,0){\makebox(0,0){$\widetilde\psi_{\rm v}$}}
\put(-28,25){\makebox(0,0){$\widetilde\psi_{\rm dv}$}}
%\put(55,20){\makebox(0,0){$\gamma_1$}}
\end{picture}
\caption{
The vertical flip $\widetilde\psi_{\rm d}$, the diagonal flip $\widetilde\psi_{\rm d}$ and their composition $\widetilde\psi_{\rm dv}=\widetilde\psi_{\rm d}\circ\widetilde\psi_{\rm v}$ on the non-singular curve $\widetilde\gamma_\lambda$.
The map $\widetilde\psi_{\rm dv}$ is the conjugate $\widetilde\pi^{-1}\circ\psi_{\rm vh}\circ\widetilde\pi$ of the map $\psi_{\rm vh}$ arising from the mutations of type $A^{(2)}_2$ with respect to $\widetilde\pi$.
}
\label{fig:diagonalflip}
\end{figure}
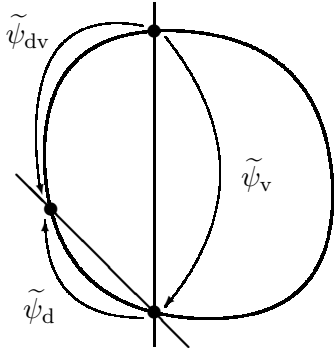

Since $\widetilde{\gamma}_\lambda$ is symmetric with respect to the line $v=u$, $\widetilde\psi_{\rm d}$ is obviously a map on $\widetilde{\gamma}_\lambda$.
Thus, by using induction on $t$,  we conclude that the composition $\widetilde\psi_{\rm dv}=\widetilde\psi_{\rm d}\circ\widetilde\psi_{\rm v}$ is a map on $\widetilde{\gamma}_\lambda-\widetilde{\mathcal{B}}$.
We call $\widetilde\psi_{\rm d}:(u^t,v^t)\mapsto(v^t,u^t)$ the diagonal flip.
Thus the map $\widetilde\psi_{\rm dv}$ is the composition $\widetilde\psi_{\rm d}\circ\widetilde\psi_{\rm v}$ of the vertical flip $\widetilde\psi_{\rm v}$ and the diagonal flip $\widetilde\psi_{\rm d}$ (see figure \ref{fig:diagonalflip}).
\qed
%---------- PROOF ----------%

%//////////////////// FIGURE ////////////////////%
%//////////////////// FIGURE ////////////////////%

We thus transform the map $\psi_{\rm vh}$ on the singular quartic curve $\gamma_\lambda$ into the map $\widetilde\psi_{\rm dv}$ on the non-singular conic $\widetilde\gamma_\lambda$.
In the subsequent sections, we give a geometric interpretation of the map $\widetilde\psi_{\rm dv}$ via a connection with the map $\phi_{\rm vh}$ arising from the mutations of type $A^{(1)}_1$.

%-------------------------------------%
%-------------- SECTION --------------%
%-------------------------------------%
\section{Commutativity of the mutations of type $\boldsymbol{A^{(2)}_2}$ and of type $\boldsymbol{A^{(1)}_1}$}
\label{sec:A11A22}
In the preceding paper \cite{Nobe16}, we investigated the birational map $\phi_{\rm vh}:\P^2(\C)\to\P^2(\C)$
\begin{align}
&\phi_{\rm vh}:(z^t,w^t)\mapsto(z^{t+1},w^{t+1}),
\nn\\
&z^{t+1}
=
\frac{(w^t)^2+1}{z^t},
\qquad
w^{t+1}
=
\frac{(z^{t+1})^2+1}{w^t}
\label{eq:QRTA11}
\end{align}
which is  arising from the mutations of type $A^{(1)}_1$ in the following manner.

Let $\mathcal{A}$ be the cluster algebra whose initial seed $(\bx_0,\by_0,B_0)$ is given by
 \begin{align*}
\bx_0=\left(x_1,x_2\right),\quad
\by_0=\left(y_1,y_2\right),\quad
B_0=\left(\begin{matrix}0&-2\\2&0\\\end{matrix}\right).
\end{align*}
We associate the variables $z^t$ and $w^t$ with the seeds of $\mathcal{A}$ as
\begin{align*}
z^t
=
\frac{x_{1;2t}}{\sqrt{y_{2;2t}}},
\qquad
w^t
=
\sqrt{y_{1;2t}}x_{2;2t}
\qquad
(t\geq0),
\end{align*}
where we define $x_{1;2t}$, $x_{2;2t}$, $y_{1;2t}$ and $y_{2;2t}$ as in (\ref{eq:mu1mu2x}) and (\ref{eq:mu1mu2y}).
All the Cartan counterparts $A(B_m)$ of the exchange matrices $B_m$ ($m\in\Z$, see (\ref{eq:mu1mu2B})) are the same and are of type $A^{(1)}_1$.
Hence, we refer to the cluster algebra $\mathcal{A}$ as of type $A^{(1)}_1$.
The mutations $\mu_1$ and $\mu_2$ are also referred as of type $A^{(1)}_1$.
The semifield $\P$ is arbitrarily chosen.
Then the exchange relation (\ref{eq:mutcv}) reduces to the map $\phi_{\rm vh}$ (see \cite{Nobe16}).

\begin{figure}[htbp]
\centering
\unitlength=.075in
\def\arraystretch{1.0}
\begin{picture}(30,5)(0,0)
\thicklines
\put(1,.5){\circle{2}}
\put(10,.5){\circle{2}}
\put(1.8,1){\vector(1,0){7.2}}
\put(1.8,0){\vector(1,0){7.2}}
\put(2.4,1){\vector(-1,0){.5}}
\put(2.4,0){\vector(-1,0){.5}}
\put(1,3){\makebox(0,0){$1$}}
\put(10,3){\makebox(0,0){$2$}}
\put(21,.5){\circle{2}}
\put(30,.5){\circle{2}}
\put(29.1,1){\vector(-1,0){7.2}}
\put(29.1,0){\vector(-1,0){7.2}}
\put(21,3){\makebox(0,0){$1$}}
\put(30,3){\makebox(0,0){$2$}}
\end{picture}
\caption{
The Dynkin diagram of type $A^{(1)}_1$ (left) and the quiver associated with the exchange matrix $B_0$ (right).
}
\label{fig:DynkinA11}
\end{figure}
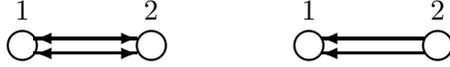

The Dynkin diagram of type $A^{(1)}_1$ and the quiver associated with the exchange matrix $B_0$ are given in figure \ref{fig:DynkinA11}.
%//////////////////// FIGURE ////////////////////%
%//////////////////// FIGURE ////////////////////%

Since $\phi_{\rm vh}$ is a QRT map \cite{QRT89}, it has an invariant curve.
We denote it by $\overline\delta_\mu$:
\begin{align*}
&\overline\delta_\mu
:=
\delta_\mu
\cup
\left\{
Q_\infty^+,
Q_\infty^-
\right\},
\\
&\delta_\mu
:=
\left(
g(z,w)
=
z^2+\mu zw+w^2+1
=
0
\right),
\end{align*}
where $\mu$ is the conserved quantity and the points $Q_\infty^+$ and $Q_\infty^-$ at infinity are given by
\begin{align*}
Q_\infty^+
=
[-\mu+\sqrt{\mu^2-4}:2:0],
\qquad
Q_\infty^-
=
[-\mu-\sqrt{\mu^2-4}:2:0]
\end{align*}
in the homogeneous coordinate $(z,w)\mapsto[z:w:1]$.
The curve $\delta_\mu$ is the affine part of $\overline\delta_\mu$.
Note that the curve $\overline\delta_\mu$ is not an elliptic curve but a conic.
The base points of the pencil $\left\{\overline\delta_\mu\right\}_{\mu\in\P^1(\C)}$ are the following four points:
\begin{align}
\left[0:\pm\sqrt{-1}:1\right],
\qquad
\left[\pm\sqrt{-1}:0:1\right].
\label{eq:BPA11}
\end{align}

%//////////////////// THEOREM ////////////////////%
%//////////////////// THEOREM ////////////////////%
%//////////////////// THEOREM ////////////////////%
\begin{remark}\label{rem:fixedness}
The points $Q_\infty^\pm$ at infinity are fixed by applying the map $\phi_{\rm vh}$.
Therefore, we consider the affine part $\delta_\mu$ as  the invariant curve of $\phi_{\rm vh}$, unless otherwise stated.
\end{remark}
%//////////////////// THEOREM ////////////////////%
%//////////////////// THEOREM ////////////////////%
%//////////////////// THEOREM ////////////////////%

%//////////////////// THEOREM ////////////////////%
%//////////////////// THEOREM ////////////////////%
%//////////////////// THEOREM ////////////////////%
\begin{remark}\label{rem:linearizability}
It is known that the map $\phi_{\rm vh}$ is linearizable.
Actually, by using the invariant curve $\delta_\mu$, we have
\begin{align*}
z^{t+1}
&=
\frac{(w^t)^2+1}{z^t}
=
\frac{-(z^t)^2-\mu z^tw^t}{z^t}
=
-z^t-\mu w^t
\\
w^{t+1}
&=
\frac{(z^{t+1})^2+1}{w^t}
=
\frac{-(w^{t+1})^2-\mu z^{t+1}w^{t+1}}{w^t}.
\end{align*}
The second equation reduces to $w^{t+1}\left(w^t+w^{t+1}+\mu z^{t+1}\right)=0$.
Since $w^{t+1}\not\equiv0$, we obtain the linearization
\begin{align*}
w^{t+1}
=
-\mu z^{t+1}-w^t
=
\mu z^t+(\mu^2-1)w^t.
\end{align*}
Here $\mu$ is given by the initial point $(z^0,w^0)$ as
\begin{align*}
\mu
=
-\frac{(z^0)^2+(w^0)^2+1}{z^0w^0}.
\end{align*}
\end{remark}
%//////////////////// THEOREM ////////////////////%
%//////////////////// THEOREM ////////////////////%
%//////////////////// THEOREM ////////////////////%

Now introduce the linear map $\omega:\P^2(\C)\to\P^2(\C)$ depending on the parameter $\lambda$:
\begin{align*}
\omega:(u,v)\mapsto(z,w)=\left(-\frac{u}{\sqrt{-\lambda}}, -\frac{v}{\sqrt{-\lambda}}\right).
\end{align*}
The map $\omega$ transforms the non-singular curve $\widetilde{\gamma}_\lambda$, which is the strict transform of the singular invariant curve $\gamma_\lambda$ of the map $\psi_{\rm vh}$ arising from the mutations of type $A^{(2)}_2$, into $\delta_\lambda$.
%//////////////////// THEOREM ////////////////////%
%//////////////////// THEOREM ////////////////////%
%//////////////////// THEOREM ////////////////////%
\begin{proposition}\label{prop:gammadelta}
If $(u,v)\in\widetilde{\gamma}_\lambda$ then $\omega(u,v)\in\delta_\lambda$, and vice versa.
\end{proposition}
%//////////////////// THEOREM ////////////////////%
%//////////////////// THEOREM ////////////////////%
%//////////////////// THEOREM ////////////////////%

%---------- PROOF ----------%
(Proof)\quad
Let $(z,w)$ be $\omega(u,v)$.
Then we have
\begin{align*}
u
=
-\sqrt{-\lambda}z,
\qquad
v=
-\sqrt{-\lambda}w.
\end{align*}
We compute
\begin{align*}
u^2+\lambda(uv-1)+v^2
=
-\lambda z^2+\lambda\left(-\lambda z w-1\right)-\lambda w^2
=
-\lambda\left( z^2+\lambda z w+ w^2+1\right).
\end{align*}
This completes the proof.
\qed
%---------- PROOF ----------%

The base points (\ref{eq:BPA11}) of the pencil $\left\{\delta_\mu\right\}_{\mu\in\C}$ are mapped into the points
\begin{align*}
\left(0,\pm\sqrt{\lambda}\right),
\qquad
\left(\pm\sqrt{\lambda},0\right)
\end{align*}
on $\widetilde{\gamma}_\lambda$ by $\omega^{-1}$, respectively.
The latter two points are the intersection points of $\widetilde\gamma_\lambda$ and the exceptional curve.
On the other hand, the set $\widetilde{\mathcal{B}}$ of the base points of $\{\widetilde\gamma_\lambda\}_{\lambda\in\C}$ is mapped into the set
\begin{align*}
\mathcal{B}_{\rm dv}
:=
\left\{
\left.
\left(
-\frac{\zeta_8^{8-i}}{\sqrt{-\lambda}}, 
-\frac{\zeta_8^i}{\sqrt{-\lambda}}
\right)
\right|
i=1,3,5,7
\right\}
\end{align*}
of points on $\delta_\lambda$ by $\omega$.

We then obtain the birational map
\begin{align*}
\phi_{\rm dv}
:=
\omega\circ\widetilde\psi_{\rm dv}\circ\omega^{-1}
=
\left(\widetilde\pi\circ\omega^{-1}\right)^{-1}\circ\psi_{\rm vh}\circ\left(\widetilde\pi\circ\omega^{-1}\right)
\end{align*}
on $\delta_\lambda-\mathcal{B}_{\rm dv}$ conjugate to the map $\widetilde\psi_{\rm dv}$ on $\widetilde{\gamma}_\lambda$ in terms of $\omega$ (see figure \ref{fig:cddelta}).
%//////////////////// FIGURE ////////////////////%
\begin{figure}[htbp]
\centering
$
\begin{CD}
\gamma_\lambda @<\widetilde\pi<< \widetilde\gamma_\lambda @> \omega >>\delta_\lambda @> \phi_{\rm vh} >>\delta_\lambda
\\
@A \psi_{\rm vh} AA @A \widetilde\psi_{\rm dv} AA @A \phi_{\rm dv} AA @VV \phi_{\rm dv} V
\\
\gamma_\lambda @<<\widetilde\pi< \widetilde\gamma_\lambda @>> \omega >\delta_\lambda @<< \phi_{\rm vh} <\delta_\lambda.
\end{CD}
$
\caption{
The commutative maps $\phi_{\rm dv}$ and $\phi_{\rm vh}$ on $\delta_\lambda$.
}
\label{fig:cddelta}
\end{figure}
%//////////////////// FIGURE ////////////////////%
The map $\phi_{\rm dv}$ is explicitly given as follows (see theorem \ref{thm:A11A22})
\begin{align}
\phi_{\rm dv}:( z^t, w^t)\mapsto
\left(\frac{( z^t)^2+1}{ w^t}, z^t\right).
\label{eq:A22ondelta}
\end{align}

The birational map $\phi_{\rm dv}$ thus obtained has a flipping structure similar to the QRT maps.
Moreover, the map $\phi_{\rm dv}$ conjugate to the mutations of type $A^{(2)}_2$ commutes with the QRT map $\phi_{\rm vh}$ arising from the mutations of type $A^{(1)}_1$.
%//////////////////// THEOREM ////////////////////%
%//////////////////// THEOREM ////////////////////%
%//////////////////// THEOREM ////////////////////%
\begin{theorem}\label{thm:A11A22}
Let $\phi_{\rm h}$ and $\phi_{\rm v}$ be the horizontal flip and the vertical flip on the curve $\delta_\lambda$, respectively:
\begin{align*}
\phi_{\rm h}:( z^t, w^t)\mapsto( z^{t+1}, w^t),
\quad
 z^{t+1}=\frac{( w^t)^2+1}{ z^t},
\\
\phi_{\rm v}:( z^t, w^t)\mapsto( z^t, w^{t+1}),
\quad
 w^{t+1}=\frac{( z^t)^2+1}{ w^t}.
\end{align*}
Also let $\phi_{\rm d}$ be the diagonal flip on $\delta_\lambda$:
\begin{align*}
\phi_{\rm d}:( z^t, w^t)\mapsto( w^t, z^t).
\end{align*}
Then we have the following decomposition of the maps $\phi_{\rm vh}$ (\ref{eq:QRTA11}) and $\phi_{\rm dv}$ (\ref{eq:A22ondelta}):
\begin{align*}
\phi_{\rm vh}=\phi_{\rm v}\circ\phi_{\rm h},
\qquad
\phi_{\rm dv}=\phi_{\rm d}\circ\phi_{\rm v}.
\end{align*}
Moreover, these maps are commutative:
\begin{align*}
\phi_{\rm vh}\circ\phi_{\rm dv}=\phi_{\rm dv}\circ\phi_{\rm vh}.
\end{align*}
\end{theorem}
%//////////////////// THEOREM ////////////////////%
%//////////////////// THEOREM ////////////////////%
%//////////////////// THEOREM ////////////////////%

%---------- PROOF ----------%
(Proof)\quad
It is clear that $\phi_{\rm vh}=\phi_{\rm v}\circ\phi_{\rm h}$ holds since $\phi_{\rm vh}$ is the QRT map.

Suppose that $(u^t,v^t)\in\widetilde{\gamma}_\lambda$, {\it viz},
\begin{align*}
\widetilde{f_0}(u^t,v^t)
=
(u^t)^2+\lambda\left(u^tv^t-1\right)+(v^t)^2
=
0.
\end{align*}
If $\omega(u^t,v^t)=( z^t, w^t)$ then we have
\begin{align*}
g( z^t, w^t)
=
( z^t)^2+\lambda z^t w^t+( w^t)^2+1
=
0
\end{align*}
by proposition \ref{prop:gammadelta}.

Now we assume that $(u^t,v^t)$ and $(u^{t+1},v^{t+1})$ satisfy $(u^{t+1},v^{t+1})=\widetilde\psi_{\rm dv}(u^t,v^t)$.
By substituting $z^t=-u^t/\sqrt{-\lambda}$ and $w^t=-v^t/\sqrt{-\lambda}$ into (\ref{eq:tildepsi}), we obtain
\begin{align*}
-\sqrt{-\lambda} z^{t+1}
&=
\frac{\sqrt{-\lambda}^3( z^t)^3-\sqrt{-\lambda} w^t}{-\lambda z^t w^t-1}.
\end{align*}
Noting that $\lambda$ is the conserved quantity:
\begin{align*}
\lambda
=
-\frac{( z^t)^2+( w^t)^2+1}{ z^t w^t},
\end{align*}
we have
\begin{align*}
 z^{t+1}
=
\frac{\lambda( z^t)^3- w^t}{\lambda z^t w^t+1}
=
\frac{\left(( z^t)^2+( w^t)^2+1\right)( z^t)^3+ z^t( w^t)^2}{\left(( z^t)^2+( w^t)^2+1\right) z^t w^t- z^t w^t}
=
\frac{( z^t)^2+1}{ w^t}.
\end{align*}

We similarly have
\begin{align*}
 w^{t+1}
&=
 z^t.
\end{align*}
Therefore, the map 
\begin{align*}
\phi_{\rm dv}: ( z^t, w^t)
\mapsto
( z^{t+1}, w^{t+1})
=
\left(
\frac{( z^t)^2+1}{ w^t},
 z^t
\right)
\end{align*}
is the composition $\phi_{\rm d}\circ\phi_{\rm v}$ of the vertical flip $\phi_{\rm d}$ and the diagonal flip $\phi_{\rm d}$.

Next we show the commutativity.
Noting that $\phi_{\rm v}$ is an involution, the condition $\phi_{\rm vh}\circ\phi_{\rm dv}=\phi_{\rm dv}\circ\phi_{\rm vh}$ to be confirmed reduces to
\begin{align*}
\phi_{\rm v}\circ\phi_{\rm h}\circ\phi_{\rm d}
=
\phi_{\rm d}\circ\phi_{\rm h}\circ\phi_{\rm v}.
\end{align*}
We then compute the LHS:
\begin{align*}
\phi_{\rm v}\circ\phi_{\rm h}\circ\phi_{\rm d}
: ( z, w)
&\overset{\phi_{\rm d}}{\mapsto}
\left( w, z\right)
\overset{\phi_{\rm h}}{\mapsto}
\left(\frac{ z^2+1}{ w}, z\right)
\overset{\phi_{\rm v}}{\mapsto}
\left(\frac{ z^2+1}{ w},\frac{\left(\frac{ z^2+1}{ w}\right)^2+1}{ z}\right)
\end{align*}
and the RHS:
\begin{align*}
\phi_{\rm d}\circ\phi_{\rm h}\circ\phi_{\rm v}
: ( z, w)
&\overset{\phi_{\rm v}}{\mapsto}
\left( z,\frac{ z^2+1}{ w}\right)
\overset{\phi_{\rm h}}{\mapsto}
\left(\frac{\left(\frac{ z^2+1}{ w}\right)^2+1}{ z},\frac{ z^2+1}{ w}\right)
\overset{\phi_{\rm d}}{\mapsto}
\left(\frac{ z^2+1}{ w},\frac{\left(\frac{ z^2+1}{ w}\right)^2+1}{ z}\right)
\end{align*}
for any $( z, w)\in\delta_\mu$ other than the base points (\ref{eq:BPA11}).
Thus, the commutativity of $\phi_{\rm vh}$ and $\phi_{\rm dv}$ is proved (see figure \ref{fig:cddelta}).
\qed
%---------- PROOF ----------%

%//////////////////// THEOREM ////////////////////%
%//////////////////// THEOREM ////////////////////%
%//////////////////// THEOREM ////////////////////%
\begin{corollary}
The map $\phi_{\rm dv}$ is linearizable:
\begin{align*}
\phi_{\rm dv}:\ ( z^t, w^t)
\mapsto
( z^{t+1}, w^{t+1})
=
\left(-\lambda z^t- w^t, z^t\right),
\end{align*}
where we put
\begin{align*}
\lambda
=
-\frac{( z^0)^2+( w^0)^2+1}{ z^0 w^0}.
\end{align*}
\end{corollary}
%//////////////////// THEOREM ////////////////////%
%//////////////////// THEOREM ////////////////////%
%//////////////////// THEOREM ////////////////////%

%---------- PROOF ----------%
(Proof)\quad
This is a direct consequence of remark \ref{rem:linearizability} and theorem \ref{thm:A11A22}.
\qed
%---------- PROOF ----------%

In the following section, we show that the maps $\phi_{\rm vh}$ and $\phi_{\rm dv}$ are endowed with their commutativity by the addition of points on an elliptic curve arising as the spectral curve of the discrete Toda lattice of type $A^{(1)}_1$.

%-------------------------------------%
%-------------- SECTION --------------%
%-------------------------------------%
\section{Geometry of the rank 2 mutations of affine types via the discrete Toda lattice of type $\boldsymbol{A^{(1)}_1}$}
\label{sec:TLCA}
In addition to the QRT map $\phi_{\rm vh}$ arising from the mutations of type $A^{(1)}_1$, we also investigated the QRT map $\varphi_{\rm vh}:\P^2(\C)\to\P^2(\C)$; 
\begin{align*}
&(x^t,y^t)\mapsto(x^{t+1},y^{t+1}),
\\
&x^{t+1}
=
\frac{-\left(bx^{t}+1\right)\left(y^{t}\right)^2}{b\left(bx^{t}+1\right)\left(y^{t}\right)^2-(a^2-b)x^{t}},
\qquad
y^{t+1}
=
\frac{\left(a^2x^{t+1}+1\right)x^{t+1}}{\left(bx^{t+1}+1\right)y^{t}}
\end{align*}
in \cite{Nobe16}.
The QRT map $\varphi_{\rm vh}$  is arising from the discrete Toda lattice of type $A^{(1)}_1$ \cite{HTI93,Suris03}, where $a,b\in\C$ are the parameters given by the initial values of the Toda lattice.

The invariant curve of the map $\varphi_{\rm vh}$ is the biquadratic curve of degree 3
\begin{align*}
&E_\nu
=
\left(
h(x,y)=0
\right)
\cup
\left\{ 
{\mathcal{O}},T,S
\right\},
\nn\\
&h(x,y):=bxy^2+y^2+\nu\left(bx^2y+xy\right)+a^2x^2+x,
\end{align*}
where the points at infinity are given by
\begin{align*}
{\mathcal{O}}:=[0:1:0],
\qquad
T:=[1:0:0],
\qquad
S:=[1:-\nu:0]
\end{align*}
in the homogeneous coordinate $(x,y)\mapsto[x:y:1]$.
The parameter $\nu$ is the conserved quantity of the QRT map $\varphi_{\rm vh}$, which is given by the initial values of the Toda lattice as well as $a$ and $b$ (see (\ref{eq:CQTLabnu})).
Note that $a$, $b$ and $\nu$ are also the conserved quantities of the discrete Toda lattice of type $A^{(1)}_1$.

The QRT map $\varphi_{\rm vh}$ and its invariant curve $E_\nu$ are respectively transformed into the discrete Toda lattice $\chi_1:\C^4\to\C^4$;
\begin{align*}
&\left(I_1^t,I_2^t,V_1^t,V_2^t\right)\mapsto\left(I_1^{t+1},I_2^{t+1},V_1^{t+1},V_2^{t+1}\right),
\\
&I_j^{t+1}
=
\frac{I_j^t+V_j^t}{I_{j+1}+V_{j+1}}I_{j+1},
\qquad
V_j^{t+1}
=
\frac{I_{j+1}^t+V_{j+1}^t}{I_{j}+V_{j}}V_{j}
\end{align*}
of type $A^{(1)}_1$ and its spectral curve as follows, where $I_j^t$ and $V_j^t$ ($j=1,2$) are the dependent variables of the Toda lattice with the subscripts reduced modulo 2.
Substitute
\begin{align*}
x=\frac{1}{v-b},
\qquad
y=
\frac{u-\nu}{v-b}
\end{align*}
into $h(x,y)=0$ we obtain
\begin{align*}
v^2+v\left(u^2-\nu u+a^2-b\right)+b\left(b-a^2\right)=0.
\end{align*}
This gives the spectral curve of the Toda lattice $\chi_1$ with imposing
\begin{align}
a^2
=
I_1^0I_2^0-V_1^0V_2^0,
\qquad
b=
-V_1^0V_2^0,
\qquad
\nu
=
-I_1^0-I_2^0-V_1^0-V_2^0,
\label{eq:CQTLabnu}
\end{align}
where $I_j^0$ and $V_j^0$ ($j=1,2$) are the initial values of the Toda lattice $\chi_1$.
Moreover, combining the map
\begin{align*}
(u,v)
\mapsto
(x,y)
=
\left(
\frac{1}{v-b},
\frac{u-\nu}{v-b}
\right)
\end{align*}
and the eigenvector map
\begin{align*}
\left(
I_1^t,I_2^t,V_1^t,V_2^t
\right)
\mapsto
(u,v)
=
\left(
-I_1^t-V_2^t,I_2^tV_2^t
\right),
\end{align*}
we obtain
\begin{align}
I_1^t
=
\frac{\left(a^2-b\right)x^t}{\left(bx^t+1\right)y^t},
\qquad
I_2^t
=
\frac{\left(bx^t+1\right)y^t}{x^t},
\qquad
V_1^t
=
-by^t,
\qquad
V_2^t
=
\frac{1}{y^t}.
\label{eq:IVxy}
\end{align}
The Toda lattice $\chi_1$ and the QRT map $\varphi_{\rm vh}$ are transformed into each other through the relation (\ref{eq:IVxy}) \cite{Nobe13,Nobe16}.

The Weierstrass model $E$ of the biquadratic curve $E_\nu$ is given as follows \cite{Silverman86,Tsuda04}
\begin{align*}
&E
=
\left(
y^2+a_1xy+a_3y-x^3-a_2x^2-a_4x-a_6=0
\right)
\cup
\left\{
\mathcal{O}
\right\},
\\
&a_1=-\nu,\quad
a_2=2a^2-b, \quad
a_3=-(a^2-b)\nu, \quad
a_4=a^2(a^2-b), \quad
a_6=0.
\end{align*}
The discriminant $\Delta$ of $E$ is
\begin{align}
\Delta
=
(a^2-b)^2 b^2 (16 a^4-8 a \nu^2+16 b \nu^2+\nu^4).
\label{eq:discriminant}
\end{align}

We see from (\ref{eq:discriminant}) that there exist two singular cases $a^2=b$ and $b=0$ for $E$. 
In both cases, the biquadratic curve $E_\nu$ is reducible, and decomposes into a line and a conic.
For $a^2=b$, the curve $E_\nu$ decomposes into
\begin{align*}
\left(a^2x+1=0\right)\cup\left\{\mathcal{O}\right\}
\qquad
\mbox{and}
\qquad
\left(y^2+\nu xy+x=0\right)\cup\left\{T,S\right\},
\end{align*}
and the map $\varphi$ on $E_\nu$ reduces to the one-dimensional one on the line $\left(a^2x+1=0\right)\cup\left\{\mathcal{O}\right\}$:
\begin{align*}
x^{t+1}
=
-\frac{1}{a^2},
\qquad
y^{t+1}
=
\frac{x^{t+1}}{y^{t}}
=
-\frac{1}{a^2y^t}.
\end{align*}
This is essentially equivalent to the mutation of type $A_1$, which is of rank 1, of finite type and has period two. 
Remark that $a_3=-(a^2-b)\nu=0$ is the necessary and sufficient condition for a non-singular QRT map to have period two \cite{Tsuda04}.

On the other hand, for $b=0$, the curve $E_\nu$ decomposes into
\begin{align*}
\left(Z=0\right)
\qquad
\mbox{and}
\qquad
\left(Y^2+\nu XY+a^2X^2+XZ=0\right),
\end{align*}
where $[X:Y:Z]=[x:y:1]$ is the homogeneous coordinate.
The map $\varphi$ reduces to the map $\phi_{\rm vh}$ arising from the mutations of type $A^{(1)}_1$, which is of rank 2 and of infinite type.
We will see the case $b=0$ later (see also \cite{Nobe16}).

We introduce an additive group structure $(E,\mathcal{O},+)$ on the Weierstrass model $E$ equipped with the unit of addition $\mathcal{O}$ in the standard manner \cite{Silverman86}.
Note that $\mathcal{O}$ is the inflection point.
The points $T$ and $S$ at infinity on $E_\nu$ correspond to
\begin{align*}
\mathcal T:=[0:0:1]
\qquad
\mbox{and}
\qquad
\mathcal S:=[0:(a^2-b)\nu:1]
\end{align*}
on $E$, respectively.
We then find
\begin{align*}
\mathcal S+\mathcal T+\mathcal{O}=\mathcal{O}
\end{align*}
because these points are on the line $X=0$.
Since $E$ and $E_\nu$ are linearly equivalent, the additive group structure on $E$ is naturally translated into the one $(E_\nu,\mathcal{O},+)$ on $E_\nu$.

It is well known that the addition of points on the Weierstrass model $E$ can be realized by using intersection of the curve and two lines.
Let $P$ and $Q$ be points on the curve $E$.
Let the line passing through both $P$ and $Q$ be $\ell_1$.
Then $\ell_1$ intersects $E$ at another point $P^\prime$.
This is a geometric realization of the algebraic relation
\begin{align*}
P+Q+P^\prime=\mathcal{O}.
\end{align*}
Let us consider the vertical line $\ell_2$ passing through the point $P^\prime$.
Then $\ell_2$ intersects $E$ at another point $P^{\prime\prime}$.
This intersection represents the algebraic relation
\begin{align*}
P^\prime+P^{\prime\prime}+\mathcal{O}=\mathcal{O}.
\end{align*}
The following algebraic relation is reduced from the above two relations
\begin{align*}
-P^\prime=P^{\prime\prime}=P+Q.
\end{align*}
Thus, the intersection point $P^{\prime\prime}$ of the curve $E$ and the line $\ell_2$ gives the addition $P+Q$ of points $P$ and $Q$.

In our biquadratic case $(E_\nu,\mathcal{O},+)$, assume that the point $Q$ is $T=[1:0:0]$. Then the line $\ell_1$ passing through both $P$ and $Q=T$ corresponds to the horizontal line $\ell_{\rm h}$ passing through $P$ since $T$ is the point at infinity. 
The horizontal line $\ell_{\rm h}$ intersects $E_\nu$ at another point $P^\prime$. (Note that $E_\nu$ is quadratic in $x$.)
Similarly,  the line $\ell_2$ corresponds to the vertical line $\ell_{\rm v}$ passing through $P^\prime$. 
The only intersection point $P^{\prime\prime}$ of $E_\nu$ and $\ell_{\rm v}$ other than $P^\prime$ gives the addition $P+T$ of points $P$ and $T$. (Note that $E_\nu$ is quadratic in $y$.)
We summarize the intersecting lines and the additions on $E_\nu$:
\begin{align*}
&\ell_{\rm h}:\
P+T+P^\prime=\mathcal{O},
\\
&\ell_{\rm v}:\ 
P^\prime+\mathcal{O}+P^{\prime\prime}=\mathcal{O}.
\end{align*}
The line $\ell_{\rm h}$ is the unique one passing through given point $P$ and $T$; and $\ell_{\rm v}$ is the unique one passing through $P^\prime$ and $\mathcal{O}$.
Since the points $T$ and $\mathcal{O}$ are the base points of the pencil $\{E_\nu\}_{\nu\in\P^1(\C)}$, the addition on $E_\nu$ realized by the intersections with $\ell_{\rm h}$ and $\ell_{\rm v}$ defines a map $\varphi_{\rm vh}$ on the pencil, uniquely.
This is a geometric interpretation of the QRT map $\varphi_{\rm vh}$ (see \cite{Tsuda04}).

%-------------------------------------%
%-------------- SUBSECTION --------------%
%-------------------------------------%
\subsection{Geometry of the mutations of type $\boldsymbol{A^{(1)}_1}$}
Employ a linear transformation $\sigma:\P^2(\C)\to\P^2(\C)$ depending on the parameter $\nu$
\begin{align*}
&\sigma:(x,y)\mapsto(p,q),
\\
&p
=
 a \epsilon\left(x-\frac{2}{\epsilon^2}\right),
\qquad
q
=
\epsilon\left(y+\frac{\nu}{\epsilon^2}\right),
\end{align*}
where we put $\epsilon:=\sqrt{\nu^2-4a^2}$.
By applying $\sigma$ to $h(x,y)$, we obtain
\begin{align*}
h\left(\frac{p}{ a \epsilon}+\frac{2}{\epsilon^2},\frac{q}{\epsilon}-\frac{\nu}{\epsilon^2}\right)
&=
\frac{1}{\epsilon^2}\left(p^2+q^2+1+\frac{\nu}{ a }pq\right)
+
\frac{b}{\epsilon^3}
\left(\frac{1}{ a }p+\frac{2}{\epsilon}\right)\left(q-\frac{\nu}{\epsilon}\right)
\left(\frac{\nu}{ a }p+q+\frac{\nu}{\epsilon}\right).
\end{align*}
Thus, the map $\sigma$ transforms the curve $E_\nu$ into $\widetilde E_\nu$ defined by
\begin{align*}
\widetilde h(x,y)
&:=
x^2+y^2+1+\frac{\nu}{ a }xy
+
\frac{b}{\epsilon}
\left(\frac{1}{ a }x+\frac{2}{\epsilon}\right)\left(y-\frac{\nu}{\epsilon}\right)
\left(\frac{\nu}{ a }x+y+\frac{\nu}{\epsilon}\right).
\end{align*}

The points $\mathcal{O}$ and $T$ at infinity on $E_\nu$ are fixed under the transformation $\sigma$, while $S$ is mapped into $[ a :-\nu:0]$ by $\sigma$.
Therefore, the horizontal flip and the vertical flip on $E_\nu$ are translated into the ones on $\widetilde E_\nu$ by $\sigma$ just as they are.

Consider the singular limit $b\to0$ (see (\ref{eq:discriminant})).
In the limit, the elliptic curve $\widetilde E_\nu$ reduces to the conic $\delta_{\nu/ a }$ which is the invariant curve of the QRT map $\phi_{\rm vh}$ arising from the mutations of type $A^{(1)}_1$.
The above observation concerning the flipping structures suggests us that the QRT map $\varphi_{\rm vh}$ simultaneously reduces to $\phi_{\rm vh}$.
Actually, we obtain the following proposition.

%//////////////////// THEOREM ////////////////////%
%//////////////////// THEOREM ////////////////////%
%//////////////////// THEOREM ////////////////////%
\begin{theorem}
Assume $b=0$.
The map $\sigma$ then transforms the QRT map $\varphi_{\rm vh}$ arising from the time evolution of the discrete Toda lattice $\chi_1$ of type $A^{(1)}_1$ into $\phi_{\rm vh}$ from the mutations of type $A^{(1)}_1$.
Namely, we have
\begin{align*}
\phi_{\rm vh}=\left.\sigma\circ\varphi_{\rm vh}\circ\sigma^{-1}\right|_{b=0}.
\end{align*}
\end{theorem}
%//////////////////// THEOREM ////////////////////%
%//////////////////// THEOREM ////////////////////%
%//////////////////// THEOREM ////////////////////%

%---------- PROOF ----------%
(Proof)\quad
Assume $b=0$.
Then $\varphi:(x^t,y^t)\mapsto(x^{t+1},y^{t+1})$ reduces to
\begin{align}
x^{t+1}
=
\frac{\left(y^{t}\right)^2}{a^2x^{t}},
\qquad
y^{t+1}
=
\frac{a^2\left(x^{t+1}\right)^2+x^{t+1}}{y^{t}}.
\label{eq:b0}
\end{align}

Consider the map $\sigma:(x^t,y^t)\mapsto(p^t,q^t)$, where
\begin{align*}
x^t
=
\frac{p^t}{ a \epsilon}+\frac{2}{\epsilon^2},
\qquad
y^t
=
\frac{q^t}{\epsilon}-\frac{\nu}{\epsilon^2}.
\end{align*}
Then the equation for $x^{t+1}$ in (\ref{eq:b0}) is transformed into
\begin{align*}
\frac{p^{t+1}}{ a \epsilon}+\frac{2}{\epsilon^2}
=
\frac{\left(\frac{q^t}{\epsilon}-\frac{\nu}{\epsilon^2}\right)^2}{a^2\left(\frac{p^t}{ a \epsilon}+\frac{2}{\epsilon^2}\right)}
=
\frac{\left(\epsilon q^t-\nu\right)^2}{\epsilon^2\left( a \epsilon p^t+2a^2\right)}.
\end{align*}
Noting $\epsilon^2=\nu^2-4a^2$, we compute
\begin{align*}
\frac{p^{t+1}}{ a }
&=
\frac{\left(\epsilon q^t-\nu\right)^2-2\left( a \epsilon p^t+2a^2\right)}{\epsilon\left( a \epsilon p^t+2a^2\right)}
=
\frac{\epsilon (q^t)^2-2\nu q^t+\epsilon-2 a p^t}{\left( a \epsilon p^t+2a^2\right)}.
\end{align*}
Since $\nu=- a \left((p^t)^2+(q^t)^2+1\right)/\left(p^tq^t\right)$, we obtain
\begin{align*}
p^{t+1}
&=
\frac{\epsilon p^t(q^t)^2+2 a \left((p^t)^2+(q^t)^2+1\right)+\epsilon p^t-2 a (p^t)^2}{\left(\epsilon p^t+2 a \right)p^t}
\\
&=
\frac{(q^t)^2\left(\epsilon p^t+2 a \right)+\epsilon p^t+2 a }{\left(\epsilon p^t+2 a \right)p^t}
\\
&=
\frac{(q^t)^2+1}{p^t}.
\end{align*}
This is nothing but the horizontal flip $\phi_{\rm h}$ composing the map $\phi_{\rm vh}=\phi_{\rm v}\circ\phi_{\rm h}$.

Similarly, we have
\begin{align*}
y^ty^{t+1}
&=
\left(\frac{q^t}{\epsilon}-\frac{\nu}{\epsilon^2}\right)\left(\frac{q^{t+1}}{\epsilon}-\frac{\nu}{\epsilon^2}\right)
\\
&=
\frac{1}{\epsilon^4}\left(\epsilon q^t-\nu\right)\left(\epsilon q^{t+1}-\nu\right)
\\
&=
\frac{1}{\epsilon^4}\left\{\epsilon^2 q^tq^{t+1}-\epsilon\nu\left(q^t+q^{t+1}\right)+\epsilon^2+4a^2\right\}.
\end{align*}
We also have
\begin{align*}
\left(a^2x^t+1\right)x^t
&=
\left( a \frac{p^t}{\epsilon}+\frac{2a}{\epsilon^2}+1\right)\left(\frac{p^t}{ a \epsilon}+\frac{2}{\epsilon^2}\right)
\\
&=
\frac{(p^t)^2}{\epsilon^2}+4 a \frac{p^t}{\epsilon^3}+\frac{4a^2}{\epsilon^4}+\frac{p^t}{ a \epsilon}+\frac{2}{\epsilon^2}
\\
&=
\frac{1}{\epsilon^4}\left(\epsilon^2(p^t)^2+4 a \epsilon p^t+4a^2+\frac{\epsilon^3p^t}{ a }+2\epsilon^2\right).
\end{align*}
Thus, the equation $y^ty^{t+1}=\left(a^2x^t+1\right)x^t$ in (\ref{eq:b0}) reduces to
\begin{align*}
\left(\epsilon q^t-\nu\right)q^{t+1}
=
\epsilon(p^t)^2+4 a  p^t+\frac{\epsilon^2p^t}{ a }+\epsilon+\nu q^t.
\end{align*}

We then compute
\begin{align*}
{\rm LHS}
&=
\frac{\epsilon p^t(q^t)^2+ a \left((p^t)^2+(q^t)^2+1\right)}{p^tq^t}\times q^{t+1}
\end{align*}
and
\begin{align*}
{\rm RHS}
&=
\epsilon(p^t)^2+\frac{ a }{p^t(q^t)^2}\left((p^t)^2+(q^t)^2+1\right)^2+\epsilon- a \frac{(p^t)^2+(q^t)^2+1}{p^t}
\\
&=
\epsilon(p^t)^2+\epsilon+ a \frac{(p^t)^2\left((p^t)^2+(q^t)^2+1\right)+(p^t)^2+(q^t)^2+1}{p^t(q^t)^2}
\\
&=
\frac{\epsilon p^t(q^t)^2+ a \left((p^t)^2+(q^t)^2+1\right)}{p^tq^t}\times\frac{(p^t)^2+1}{q^t}.
\end{align*}
Therefore, we obtain
\begin{align*}
q^{t+1}
=
\frac{(p^t)^2+1}{q^t}.
\end{align*}
This is the vertical flip $\phi_{\rm v}$ composing the map $\phi_{\rm vh}=\phi_{\rm v}\circ\phi_{\rm h}$.
\qed
%---------- PROOF ----------%

%//////////////////// THEOREM ////////////////////%
%//////////////////// THEOREM ////////////////////%
%//////////////////// THEOREM ////////////////////%
\begin{remark}\label{rem:precedingpaper}
In \cite{Nobe16}, we apply another transformation $\rho:(z,w)\mapsto(x,y)=(z^2,zw)$ to the invariant curve $\delta_\nu$ of the map $\phi_{\rm vh}$ arising from the mutations of type $A^{(1)}_1$, and obtain the invariant curve $E_\nu$ (with imposing $b=0$) of the map $\varphi_{\rm vh}$ arising from the discrete Toda lattice $\chi_1$ of type $A^{(1)}_1$.
Through $\rho$, we relate the cluster variables with the solution to the Toda lattice for a special choice of the initial values.
Since $\rho$ is independent of the parameter $\nu$ differently from the one $\sigma$ employed above, $\rho$ maps the pencil $\{\delta_\nu\}_{\nu\in\P^1(\C)}$ into the pencil $\{E_\nu\}_{\nu\in\P^1(\C)}$ all at once; however, it is not birational but rational.
In order to clarify geometry of the mutations of affine types in terms of the additive group structure of the elliptic curve $E_\nu$, we employ the birational transformation $\sigma$ depending on the parameter $\nu$ between $E_\nu$ and $\widetilde E_\nu$, which reduces to $\delta_{\nu/a}$ in the limit $b\to0$.
\end{remark}
%//////////////////// THEOREM ////////////////////%
%//////////////////// THEOREM ////////////////////%
%//////////////////// THEOREM ////////////////////%

%-------------------------------------%
%-------------- SUBSECTION --------------%
%-------------------------------------%
\subsection{Geometry of the mutations of type $\boldsymbol{A^{(2)}_2}$}
In this subsection, we will show that, in the singular limit $b\to0$, a certain point addition on the biquadratic curve $E_\nu$ reduces to the map $\phi_{\rm dv}$ on the conic $\delta_\lambda$ arising from the mutations of type $A^{(2)}_2$, similar to the case of the mutations of type $A^{(1)}_1$ discussed above.
Since the flipping structures of the mutations of type $A^{(1)}_1$ has already shown, we have only to show that the diagonal flip on $\delta_\lambda$ can be derived from a point addition on $E_\nu$.

Let $P=(p_1,p_2)$ be a point on the affine part of $E_\nu$.
Then $-P=(p_1,\overline p_2)$ is the unique intersection point of $E_\nu$ and $\ell_{\rm v}$, the vertical line $x=p_1$ passing through $P$, where
\begin{align*}
\overline p_2
=
\frac{\left(a^2p_1+1\right)p_1}{\left(bp_1+1\right)p_2}.
\end{align*}
Consider the line $\ell_{\rm d}$ passing through $-P$ with slope $- a $:
\begin{align*}
 a (x-p_1)+y-\overline p_2=0.
\end{align*}
Then $\ell_{\rm d}$ intersects $E_\nu$ at two points $Q=(q_1,q_2)$ and $R=(r_1,r_2)$ other than $-P$, where $x=q_1,r_1$ solve the equation
\begin{align}
( a -\nu) a bx^2
+
\left\{(2 a -\nu)( a -b\overline p_2)+abp_1\right\}x
-
\frac{( a p_1+\overline p_2)^2}{p_1}
=
0
\label{eq:L4Eint}
\end{align}
and are expanded by $b$ around $b=0$ as follows
\begin{align*}
&q_1
=
\frac{\left( a p_1+p_2\right)^2}{ a \left(2 a -\nu\right)p_1}
+
O(b),
\\
&r_1
=
-\frac{2 a -\nu}{\left( a -\nu\right)b}+O(1).
\end{align*}
The equation (\ref{eq:L4Eint}) follows by eliminating $y$ from the equations of $\ell_{\rm d}$ and $E_\nu$.
The intersection of $\ell_{\rm d}$ and $E_\nu$ is a geometric realization of the algebraic relation
\begin{align*}
-P+Q+R=\mathcal{O}.
\end{align*}
Hence, $Q$ is equivalent to the addition $P-R$.
We denote the map which maps $P$ into $P-R$ by $\varphi_{\rm dv}$, since it is the composition of the vertical flip $P\mapsto-P$ given by $\ell_{\rm v}$ and the diagonal flip $-P\mapsto P-R$ by $\ell_{\rm d}$.

The discrete Toda lattice $\varphi_{\rm vh}$ of type $A^{(1)}_1$ is realized as the point addition $P\mapsto P+T$ given by the intersection $E_\nu$, $\ell_{\rm h}$ and $\ell_{\rm v}$.
Also, the map $\varphi_{\rm dv}$ is realized as the point addition $P\mapsto P-R$ given by the intersection of $E_\nu$,  $\ell_{\rm v}$ and $\ell_{\rm d}$.
Therefore, these maps $\varphi_{\rm vh}$ and $\varphi_{\rm dv}$ are commutative since the additive group $(E_\nu,\mathcal{O},+)$ is abelian.
Thus, the map $\varphi_{\rm dv}$ is the commuting flow of the discrete Toda lattice  $\varphi_{\rm vh}$ of type $A^{(1)}_1$, namely,  the B\"acklund transformation of it.

Now we show that the map $\varphi_{\rm dv}$ on $\widetilde E_\nu$ reduces to the map $\phi_{\rm dv}$ on $\delta_{\nu/a}$, which is arising from the mutations of type $A^{(2)}_2$, in the limit $b\to0$.
Consider the limit $b\to0$ of $q_1$, $q_2$ and $r_1$:
\begin{align*}
&\lim_{b\to0}q_1
=
\frac{\left( a p_1+p_2\right)^2}{ a \left(2 a -\nu\right)p_1}
=
\frac{1}{ a }\left(p_2-\frac{1}{2 a -\nu}\right)
=:\widetilde q_1,
\\
&\lim_{b\to0}q_2
=
 a p_1+\frac{1}{2 a -\nu}
=:\widetilde q_2,
\\
&\lim_{b\to0}r_1
=
\infty.
\end{align*}
Notice that we have 
\begin{align*}
\nu
=
-\frac{a^2p^2+p+q^2}{pq}
\end{align*}
in the limit $b\to0$.

Let
\begin{align*}
\widetilde p_1:=\lim_{b\to0}p_1,
\qquad
\widetilde p_2:=\lim_{b\to0}\overline p_2.
\end{align*}
Put $(\xi_1,\xi_2)=\sigma(\widetilde p_1,\widetilde p_2)$ and $(\eta_1,\eta_2)=\sigma(\widetilde q_1,\widetilde q_2)$.
We then have
\begin{align*}
\eta_1
&=
 a \epsilon\left(\widetilde q_1-\frac{2}{\epsilon^2}\right)
=
 a \epsilon\left\{\frac{1}{ a }\left(q-\frac{1}{2 a -\nu}\right)-\frac{2}{\epsilon^2}\right\}
\\
&=
\epsilon q-\frac{\epsilon^2+4a^4-2 a \nu}{\left(2 a -\nu\right)\epsilon}
\\
&=
\epsilon q+\frac{\nu}{\epsilon}
=
\xi_2
\end{align*}
and
\begin{align*}
\eta_2
&=
\epsilon\left(\widetilde q_2+\frac{\nu}{\epsilon^2}\right)
=
\epsilon\left( a p+\frac{1}{2 a -\nu}+\frac{\nu}{\epsilon^2}\right)
\\
&=
 a \epsilon p+\frac{\epsilon^2+ a \nu-\nu^2}{\left(2 a -\nu\right)\epsilon}
\\
&=
 a \epsilon p-\frac{2 a }{\epsilon}
=
\xi_1.
\end{align*}

Thus, in the limit $b\to0$, the map $\varphi_{\rm dv}$ on $\widetilde E_\nu$  which maps $P$ into $Q=P-R$ reduces to the composition $\phi_{\rm d}\circ\phi_{\rm v}$ of the vertical flip $\phi_{\rm v}$ and the diagonal flip $\phi_{\rm d}$ on $\delta_{\nu/ a }$.
This is nothing but the map $\phi_{\rm dv}$ on $\delta_{\lambda}$ with $\lambda=\nu/ a $ arising from the mutations of type $A^{(2)}_2$.
Thus, we conclude that the map $\phi_{\rm dv}$ is reduced as the singular limit $b\to0$ of the addition $Q=P-R$ on the biquadratic curve $E_\nu$.
We complete this in the following theorem.

\begin{figure}[htb]
\centering
\unitlength=.03in
\def\arraystretch{1.0}
\begin{picture}(100,100)(-25,-20)
\put(-40,-22){\dashbox{.5}(13, 37){}}
\put(53,27){\dashbox{.5}(45, 35){}}
\put(3,-22){\dashbox{.5}(45, 37){}}
\thicklines
\put(-32,22){\makebox(0,0){singular curve}}
\put(25,22){\makebox(0,0){conic}}
\put(75,68){\makebox(0,0){elliptic curve}}
\put(28,20){\makebox(0,0){
$
\xymatrix@C=36pt@R=36pt{
   &&&& E_\nu \ar[r]^{\varphi_{\rm vh}} \ar@/_40pt/[lldd]_{\sigma|_{b=0}} & E_\nu \ar[d]^{\varphi_{\rm dv}}  \ar@/_15pt/[lldd]_(.7){\sigma|_{b=0}} |(.42)\hole
    \\
   &&&& E_\nu \ar[r]_{\varphi_{\rm vh}} \ar[u]^{\varphi_{\rm dv}} \ar@/^15pt/[lldd]^(.25){\sigma|_{b=0}} |(.61)\hole & E_\nu \ar@/^40pt/[lldd]^{\sigma|_{b=0}}
    \\
    \gamma_\lambda && \delta_\lambda \ar[ll]_{\widetilde\pi\circ\omega^{-1}} \ar[r]^{\phi_{\rm vh}} & \delta_\lambda \ar[d]^{\phi_{\rm dv}} 
    \\
    \gamma_\lambda \ar[u]^{\psi_{\rm vh}} && \delta_\lambda \ar[ll]^{\widetilde\pi\circ\omega^{-1}} \ar[r]_{\phi_{\rm vh}} \ar[u]^{\phi_{\rm dv}} & \delta_\lambda
    }
$}}
\end{picture}
\caption{
Correspondence among the maps $\psi_{\rm vh}$ on the singular curve $\gamma_\lambda$ arising from the mutations of type $A^{(2)}_2$, $\phi_{\rm dv}=\left(\widetilde\pi\circ\omega^{-1}\right)^{-1}\circ\psi_{\rm vh}\circ\left(\widetilde\pi\circ\omega^{-1}\right)$ on the conic $\delta_\lambda$, $\phi_{\rm vh}$ on $\delta_\lambda$ from the mutations of type $A^{(1)}_1$, $\varphi_{\rm vh}$ on the elliptic curve $E_\nu$ from the time evolution of the discrete Toda lattice $\chi_1$ of type $A^{(1)}_1$ and $\varphi_{\rm dv}$ on $E_\nu$ from the B\"acklund transformation of the Toda lattice.
}
\label{fig:relationsofmaps}
\end{figure}
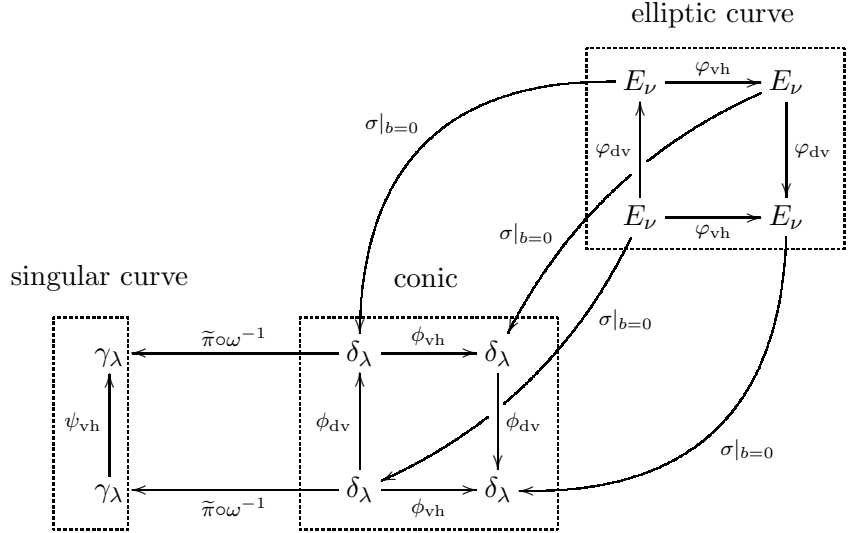

%//////////////////// THEOREM ////////////////////%
%//////////////////// THEOREM ////////////////////%
%//////////////////// THEOREM ////////////////////%
\begin{theorem}
Assume $b=0$.
The map $\sigma$ then transforms the  map $\varphi_{\rm dv}$ arising from the B\"acklund transformation of the discrete Toda lattice $\chi_1$ of type $A^{(1)}_1$ into $\phi_{\rm dv}$ from the mutations of type $A^{(2)}_2$.
Namely, we have
\begin{align*}
\phi_{\rm dv}=\left.\sigma\circ\varphi_{\rm dv}\circ\sigma^{-1}\right|_{b=0}
\end{align*}
(see figure \ref{fig:relationsofmaps}).\qed
\end{theorem}
%//////////////////// THEOREM ////////////////////%
%//////////////////// THEOREM ////////////////////%
%//////////////////// THEOREM ////////////////////%

Finally, we summarize the correspondence among the maps $\psi_{\rm vh}$, $\phi_{\rm dv}$ arising from the mutations of type $A^{(2)}_2$, $\phi_{\rm vh}$ from the mutations of type $A^{(1)}_1$ and $\varphi_{\rm vh}$, $\varphi_{\rm dv}$ from the discrete Toda lattice $\chi_1$ of type $A^{(1)}_1$ in figure \ref{fig:relationsofmaps}.
Commutativity of the maps $\varphi_{\rm vh}$ and $\varphi_{\rm dv}$ on the elliptic curve $E_\nu$ reduces to that of the maps $\phi_{\rm vh}$ and $\phi_{\rm dv}$ on the conic $\delta_\lambda$.
%//////////////////// FIGURE ////////////////////%
%//////////////////// FIGURE ////////////////////%

%-------------------------------------%
%-------------- SECTION --------------%
%-------------------------------------%
\section{Concluding remarks}
\label{sec:CONCL}
We constructed the discrete integrable system on the singular quartic curve $\overline\gamma_\lambda$ governed by the quartic birational map $\psi_{\rm vh}$ from the mutations of type $A^{(2)}_2$.
The singularity of the curve $\overline\gamma_\lambda$ was then resolved by blowing-up the curve at the singular point $\mathfrak{p}$, and the non-singular curve birationally equivalent to the conic $\delta_\lambda$ was obtained.
The birational map was simultaneously transformed into the one $\phi_{\rm dv}$ on the conic $\delta_\lambda$ by the blowing-up.
The conic $\delta_\lambda$ thus obtained is nothing but the invariant curve of the birational map $\phi_{\rm vh}$ arising from the mutations of type $A^{(1)}_1$.
Moreover, these two birational maps $\phi_{\rm dv}$ and $\phi_{\rm vh}$ are commutative on the conic $\delta_\lambda$ since they have flipping structures given by the intersection of $\delta_\lambda$ and several lines.
We finally showed that the flipping structures came from the additive group structure on the elliptic curve $E_\nu$ arising as the spectral curve of the discrete Toda lattice $\chi_1$ of type $A^{(1)}_1$.
It follows that commutativity of the time evolution and the B\"acklund transformation of the Toda lattice on the elliptic curve $E_\nu$ reduces to commutativity of the bitrational maps $\phi_{\rm dv}$ and $\phi_{\rm vh}$ on the conic $\delta_\lambda$ in the singular limit $b\to0$.

In this paper, we revealed integrable structures of the rank 2 mutations of affine types from the viewpoint of addition of points on the elliptic curve.
In the forthcoming paper \cite{Nobe18}, we will investigate the rank 2 mutations of finite type, namely, of types $A_1\times A_1$, $A_2$, $B_2$ and $G_2$.
Although we have already obtained the birational maps and the invariant curves arising from these mutations in the preceding papaer \cite{Nobe16}, their geometries have not revealed precisely, yet.
We will present a new invariant curve of the birational map arising from the mutations of type $G_2$, which is a quartic singular curve.
(Note that the mutations of finite type have a certain finite period, therefore, the map arising from it has several invariant curves.)
Resolution of the singularity of the curve gives a geometric interpretation of the map arising from the mutations of type $G_2$ in terms of addition of points on an elliptic curve.
By using additive group structures on elliptic curves, we will complete classification of the rank 2 cluster algebras of finite and affine types.

\section*{Acknowledgments}
This work was partially supported by JSPS KAKENHI Grant Number 26400107.

%######################################################%
%#################### BIBLIOGRAPHY ####################%
%######################################################%
%\section*{References}

\end{document}